\documentstyle[epsf,graphics,epsfig]{mn}

\def\spose#1{\hbox to 0pt{#1\hss}}
\def\simlt{\mathrel{\spose{\lower 3pt\hbox{$\mathchar"218$}}
     \raise 2.0pt\hbox{$\mathchar"13C$}}}
\def\simgt{\mathrel{\spose{\lower 3pt\hbox{$\mathchar"218$}}
     \raise 2.0pt\hbox{$\mathchar"13E$}}}
\newcommand{\etal}{{\it et~al.}}

\title[Quasar host galaxies]
{The host galaxies of luminous radio-quiet quasars}

\author[W.J. Percival \etal]{W.J.Percival,$^{1,2}$ L.Miller,$^1$
	R.J. McLure$^{2,1}$ and J.S. Dunlop$^2$\\
  $^1$ Dept. of Physics, University of Oxford, 
  Nuclear \& Astrophysics Laboratory, Keble Road, Oxford OX1 3RH, U.K.\\
  $^2$ Institute for Astronomy, University of Edinburgh, 
  Royal Observatory, Blackford Hill, Edinburgh EH9 3HJ, U.K.\\}

\date{Submitted for publication in MNRAS}
\begin{document}
\maketitle

\begin{abstract} 
We present the results of a deep $K$-band imaging study which reveals
the host galaxies around a sample of luminous radio-quiet quasars. The
$K$-band images, obtained at UKIRT, are of sufficient quality to allow
accurate modelling of the underlying host galaxy. Initially, the basic
structure of the hosts is revealed using a modified Clean
deconvolution routine optimised for this analysis. 2 of the 14 quasars
are shown to have host galaxies with violently disturbed morphologies
which cannot be modelled by smooth elliptical profiles. For the
remainder of our sample, 2D models of the host and nuclear component
are fitted to the images using the $\chi^{2}$ statistic to determine
goodness of fit. Host galaxies are detected around all of the
quasars. The reliability of the modelling is extensively tested, and
we find the host luminosity to be well constrained for 9 quasars. The
derived average $K$-band absolute $K$-corrected host galaxy magnitude
for these luminous radio-quiet quasars is $\langle M_K
\rangle=-25.15\pm0.04$, slightly more luminous than an $L^*$
galaxy. The spread of derived host galaxy luminosities is small,
although the spread of nuclear-to-host ratios is not. These host
luminosities are shown to be comparable to those derived from samples
of quasars of lower total luminosity and we conclude that there is no
correlation between host and nuclear luminosity for these
quasars. Nuclear-to-host ratios break the lower limit previously
suggested from studies of lower nuclear luminosity quasars and Seyfert
galaxies. Morphologies are less certain but, on the scales probed by
these images, some hosts appear to be dominated by spheroids but
others appear to have disk-dominated profiles.
\end{abstract}

\begin{keywords}
galaxies: active, quasars: general, infrared: galaxies
\end{keywords}

\section{Introduction}
Models of the cosmological evolution of quasars often use galaxy
mergers as the primary mechanism for quasar activation and require the
mass of the structure within which a quasar is formed as a basic
parameter \cite{efrees,haehnelt,percival}. One step towards testing
hypotheses about quasar initiation is to answer the question: Is a
quasar's luminosity correlated with the luminosity of the structure
within which it formed? Such a correlation has been shown to exist for
low redshift ($0<z<0.3$) Seyferts and quasars with luminosities
$M_V\simgt-25$ in that there appears to be a lower limit to the host
luminosity which increases with quasar luminosity
\cite{mcleod95a,mcleod99}. However this limit is poorly defined,
particularly for high luminosity quasars when the strong nuclear
component makes it increasingly more difficult to find low luminosity
hosts.

Recent work has shown that the majority of nearby galaxies have
massive dark objects in their cores, which are suggested to be
super-massive black holes potentially capable of powering AGN
\cite{kormendy,magorrian}. These studies have also found evidence for
a correlation between the mass of the compact object and the
luminosity of the spheroidal component of the host. Assuming a link
between nuclear luminosity and black hole mass, the average nuclear
luminosity emitted by low redshift quasars is expected to increase
with host spheroidal luminosity. In light of this prediction there has
been a resurgence of interest in host galaxy studies and recent work
\cite{mclure} has found weak evidence for a correlation in accord with
the relations of Magorrian \etal\ \shortcite{magorrian}. However, this
correlation relies on spheroid/disk decomposition for two quasars with
low nuclear luminosities, and only a small number of luminous
radio-quiet quasars were observed.

Host galaxy properties of AGN are known to be correlated with the
radio power: radio galaxies tend to be large spheroidal galaxies,
while disk galaxies tend to be radio-quiet. Recent evidence suggests
that the hosts of radio-loud quasars are also predominantly massive
spheroidal galaxies regardless of the nuclear luminosity
\cite{dunlop93,taylor,mclure}. However, studies of radio-quiet quasars
with luminosities $M_V\simgt-25$ have shown that the hosts can be
dominated by either disk-like or spheroidal components or can be
complex systems of gravitationally interacting components
\cite{taylor,bahcall,boyce,mclure}. There is therefore strong
justification for studies to see if these luminosity and morphological
trends extend to the hosts of more luminous ($M_V\simlt-25$)
radio-quiet quasars.

There have been many recent detections of host galaxies in the optical
thanks to results from HST \cite{hooper,bahcall,boyce,mclure}, which
add to our knowledge from ground-based studies
\cite{malkan84a,malkan84b,hutchings92,veron-cetty}. However, quasar
hosts often appear significantly disturbed, as if by interaction or
merger which can lead to strong bursts of star-formation and
significant extended line and blue continuum emission at optical
wavelengths which are not indicative of the mass of the underlying
host. The nuclear-to-host light ratio in the optical is also typically
higher than at longer wavebands.

These problems can be circumvented by observing in the infrared where
the contrast between host and nuclear component is improved and the
emission associated with star bursting activity is largely absent: the
$K$ magnitude is a better measurement of the long-lived stellar
populations in the host \cite{bruzual}. Previous observations in the
infra-red have been successfully used to determine quasar host galaxy
luminosities and morphologies (McLeod \& Rieke 1994a;b; Dunlop \etal\
1993; Taylor \etal\ 1996). However, recent advances in telescope
design, in particular the advent of adaptive optics systems such as
the tip-tilt system on UKIRT produce clearer images of quasars and
enable accurate point spread functions (psfs) to be more readily
obtained as differences between successive observations are
reduced. Such advances coupled with improved analysis techniques mean
we are now able to reveal the host galaxies of luminous quasars with
$M_V\simlt-25$ using infra-red observations.

In order to obtain enough luminous radio-quiet quasars our sample was
forced to cover redshifts $0.26\le z\le0.46$. At such redshifts, with
typical seeing, the structure of the host galaxy is hidden in the
wings of the psf from the nuclear component. There are two main ways
of proceeding: either the psf can be deconvolved from the quasar light
to directly observe the host galaxy, or known galaxy profiles can be
used to model the hosts, a nuclear component can be added in and the
profiles can be fitted to the data. Because the host galaxies
sometimes have disturbed morphologies indicative of violent mergers,
it is difficult to assume a form for the galaxy. However without such
modelling, it is not easy to determine the contribution of the host to
the light from the centre of the quasar and deconvolution routines
tend to produce biased solutions which may alter important features.

For the analysis of our quasar sample, an approach is adopted which
uses both methods. Initially (Section~\ref{sec:simple}) the images
were restored using a deconvolution algorithm, based on the Clean
algorithm \cite{hogbom}, developed for this problem, which will be
described elsewhere \cite{percival_clean}. This routine was used to
reveal the extent to which the `nebulosity' around the point source is
disturbed. Deconvolution of the light from two of our quasars reveals
violently disturbed host galaxies indicative of close merger
events. In the remainder of our sample, the non-nuclear light is more
uniformly distributed around the centre of the quasar. We should note
that the resolution provided by this deconvolution technique is
probably not sufficient to reveal evidence for weak mergers, where the
host galaxy is only slightly disturbed.

Where the image-restoration routine revealed approximate elliptical
symmetry in the non-nuclear component, 2D galaxy profiles were fitted
to the hosts. Analysis of non-interacting, low redshift galaxies has
shown that an empirical fit to both disk and spheroidal systems is
given by:
\begin{equation}
  \mu=\mu_o\exp\left[-\left(\frac{r}{r_o}\right)^{1/\beta}\right].
  \label{eq:galprofile}
\end{equation}
where $\mu$ is the average surface brightness in concentric elliptical
annuli around the core, and $r$ is the geometric average of the
semi-major and semi-minor axes.

Model images were carefully created using this profile and were tested
against the data using the $\chi^2$ statistic to determine goodness of
fit. Five host parameters were required, the half-light radius,
integrated luminosity, axial ratio, angle on the sky, and the
power-law parameter of the galaxy $\beta$, as well as the
nuclear-to-host ratio. Section~\ref{sec:model} describes the modelling
procedure in detail, and in Section~\ref{sec:results} the best-fit
parameters are presented for the host galaxies.

Much previous work has produced ambiguous results because of a lack of
error analysis and insufficient testing of the modelling. A detailed
analysis of the reliability of the 2D modelling method used in this
paper has therefore been undertaken and is presented in
Section~\ref{sec:test}. Although hosts are detected in all of our
sample, the upper limit of the host luminosity is only usefully
constrained for 9 of the 12 quasars modelled (see
Section~\ref{sec:calc_lum}). Similar analysis of the best-fit $\beta$
parameter which determines the morphology of the host reveals that
this parameter is, unsurprisingly, more poorly constrained than the
luminosity. However, we have created Monte-Carlo simulations of images
with the same signal-to-noise as the original images
(Sections~\ref{sec:mock}~\&~\ref{sec:mock2}). By analysing these
images using exactly the same procedure as for the original data we
find that it is possible to distinguish between disk and spheroidal
structure.

Unless stated otherwise we have adopted a flat, $\Lambda=0$
cosmological model with $H_0=50$km\,s$^{-1}$Mpc$^{-1}$ and have
converted previously published data to this cosmology for ease of
comparison.

\section{The Sample and Observations}

\begin{table*}
\begin{minipage}{\textwidth}
  \centering
  \begin{tabular}{ccccccccclc} \hline
    quasar & \multicolumn{2}{c}{J2000 coords} & V & z & $M_V$ 
      & 1.4\,GHz Flux density & Observing run &
      \multicolumn{3}{c}{Integration time} \\ 
    & & & & & & /W\,Hz$^{-1}$Sr$^{-1}$ & & & /s & \\ \hline 
    0043+039 & 00 45 47.3 & +04 10 22.5 & 16.0 & 0.384 & $-$26.0 & 
        -                 & 09/1997 & & 2800 & \\
    0137$-$010 & 01 40 17.0 & $-$00 50 03.0 & 16.4 & 0.335 & $-$25.3 & 
      1.46$\times10^{23}$ & 09/1996 & & 11300 & \\
    0244$-$012 & 02 46 51.8 & $-$00 59 32.3 & 16.5 & 0.467 & $-$25.9 & 
        -                 & 09/1997 & & 10300 & \\
    0316$-$346 & 03 18 06.5 & $-$34 26 37.1 & 15.1 & 0.265 & $-$26.0 & 
        -                 & 09/1996 & & 6400 & \\
    0956$-$073 & 09 59 16.7 & $-$07 35 18.9 & 16.5 & 0.327 & $-$25.1 & 
        -                 & 05/1998 & & 8000 & \\
    1214+180 & 12 16 49.1 & +17 48 04.1 & 16.7 & 0.374 & $-$25.2 & 
        -                 & 05/1998 & & 7590 & \\
    1216+069 & 12 19 20.9 & +06 38 38.4 & 15.7 & 0.334 & $-$26.0 & 
        -                 & 05/1998 & & 7200 & \\
    1354+213 & 13 56 32.9 & +21 03 51.2 & 15.9 & 0.300 & $-$25.5 & 
        -                 & 05/1998 & & 8600 & \\
    1543+489 & 15 45 30.2 & +48 46 08.9 & 16.1 & 0.400 & $-$26.0 & 
        -                 & 05/1998 & & 14300 & \\
    1636+384 & 16 38 17.6 & +38 22 49.0 & 17.0 & 0.360 & $-$24.8 & 
        -                 & 05/1998 & & 7100 & \\
    1700+518 & 17 01 24.9 & +51 49 20.4 & 15.1 & 0.290 & $-$26.2 & 
      7.18$\times10^{23}$ & 09/1997 & & 3600 & \\
    2112+059 & 21 14 52.6 & +06 07 42.5 & 15.5 & 0.466 & $-$26.7 & 
      2.95$\times10^{23}$ & 09/1997 & & 15100 & \\
    2233+134 & 22 36 07.7 & +13 43 55.0 & 16.7 & 0.325 & $-$26.9 & 
        -                 & 09/1996 & & 6400 & \\
    2245+004 & 22 47 41.6 & +00 54 57.3 & 18.5 & 0.364 & $-$23.4 & 
        -                 & 09/1997 & & 11100 & \\ \hline
  \end{tabular}

  \caption{The sample of 14 quasars observed. $M_V$ was calculated for
  each quasar from apparent magnitudes given in the catalogue of
  Hewitt \& Burbidge \protect\shortcite{hewitt} assuming no
  $K$-correction. Radio fluxes were determined using the NVSS 1.4\,GHz
  survey. Only 3 of the quasars were detected in this survey and their
  radio fluxes are below the radio-quiet/loud cutoff. We observed this
  sample in 3 observing runs at UKIRT. For the 09/1996 run, the
  tip-tilt system was not operational and the psf stars are not
  expected to be as well matched to the quasars as for the other runs
  (see Section~\protect\ref{sec:psf}).  Note that coordinates given
  were determined using the original finding charts and the Digitised
  Sky Survey and may be different from those in the catalogue of
  Hewitt \& Burbidge \protect\shortcite{hewitt} which are often
  inaccurate.}  \label{tab:quasars}
\end{minipage}
\end{table*}

We have selected 13 luminous ($M_V\le-25.0$) quasars and one less
luminous quasar within the redshift range $0.26\le z\le0.46$. The
quasars were checked for radio loudness using the NVSS survey
\cite{condon}. Three of the 14 quasars were detected at 1.4\,GHz in
this survey (see Table~\ref{tab:quasars}) but their flux densities are
all $<10^{24}$\,W\,Hz$^{-1}$Sr$^{-1}$ and they are considered part of
the radio-quiet population. Three quasars, 0137$-$010, 0316$-$346 and
2233+134 were observed at UKIRT before the tip-tilt system was
operational and so these data are not of the same quality as those
from subsequent runs.

Of the 13 luminous quasars selected, three, 0956$-$073, 1214+186 and
1636+384 have not had previous attempts to measure host magnitudes and
morphologies. It is difficult to assess the significance of claimed
host detections for the other quasars and the associated parameters
calculated because of the lack of error analysis which abounds in this
field, and the great potential for systematic errors caused by the
requirement for accurate psf measurements. However, individual results
from these studies are compared to the results of this paper in
Section~\ref{sec:results-quasars}.

The observations were all taken using the $256\times256$\,pixel InSb
array camera IRCAM\,3 on the 3.9\,m UK Infrared Telescope (UKIRT). The
pixel scale is 0.281\,arcsec\,pixel$^{-1}$ which gives a field of view
of $\sim$72\,arcsec. Our sample of quasars was observed during three
observing runs in 09/1996, 09/1997 and 05/1998. For the later two runs
the image quality was exceptional with consistent FWHM of 0.45\,arcsec
observed. 

The $K$-band quasar images were taken using a quadrant jitter pattern.
This cycled 2 or 4 times through a 4-point mosaic placing the quasar
in each of the quadrants in turn. The actual position of the central
value within each quadrant was shifted slightly for each image to
reduce the effect of bad pixels. Each image consists of
$\sim$100\,secs of integration time divided into exposures calculated
to avoid saturation. The exposures varied between 5-10\,secs for the
quasars alone, down to 0.2\,secs for the quasars with a bright star on
the chip which we hoped to use as a psf star. Standard stars from the
sample of UKIRT faint standards \cite{casali} were observed for
photometric calibration between observations of different quasars. All
of the images were corrected for the non-linear response of IRCAM\,3
using a formula supplied by the telescope support staff.

\section{Obtaining the correct PSF} \label{sec:psf}

Obtaining an accurate psf is vital to the analysis of the images. With
these ground-based observations the psf varies with seeing conditions
and telescope pointing. An experimental psf was therefore determined
for each of the quasars by observing a nearby bright star. This led to
an unbiased, accurate psf without recourse to the quasar images. For
three of the quasars, 0956$-$073, 1214+180 and 1216+069 there was a
nearby star which could be placed on the frame with the quasar. This
gave an accurate psf measurement with no loss of integration time on
the quasar. If required, the position of the quasar for each
observation was altered slightly to allow both the quasar and psf to
be well within the boundaries of the chip. For the remaining quasars
the telescope was offset to a nearby bright star to use as the psf,
before and after each quasar integration (which lasted a maximum of
1600\,secs). A number of psf measurements were therefore obtained for
each night and each quasar. To ensure consistent adaptive optics
correction, properties of the tip-tilt guiding were matched between
quasar and psf measurements. To do this psf stars were selected to
enable tip-tilt guiding from a star of a similar magnitude, distance
from the object, and position angle to that used for the quasar
image. Magnitudes of the stars chosen to provide a psf measurement are
given in Table~\ref{tab:chisq}. By examining fine resolution contour
plots of the psf images, it was found that the psf was stable over the
course of each night, but varied between nights at the telescope and
for different telescope pointing. Because of this, the final stacking
of psf images was performed with the same weighting between days as
for the quasar images (see Section~\ref{sec:data}).

\begin{figure}
  \centering
  \resizebox{\columnwidth}{!}{\includegraphics{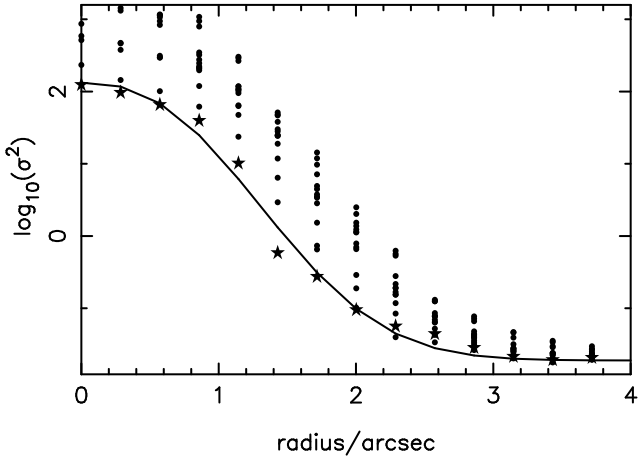}}
  \caption{ The radial profile of the variance (measured in annuli of
  width 0.5\,arcsec) of the difference between the image of quasar
  1543+489 and the scaled psf obtained using the procedure in
  Section~\protect\ref{sec:psf} (stars). For comparison, the profiles
  of the variance obtained using the 13 other psf measurements are
  also plotted (solid circles). The solid line is the best fit model
  to the variance, calculated as in Section~\protect\ref{sec:error}.}
\label{fig:psfquality}
\end{figure}

As a test of the effectiveness of this procedure to provide the
correct psf, the fit between measured psf and image for quasar
1543+489 has been compared to the fit between psfs measured for
different quasars and the same image.  Fig.~\ref{fig:psfquality} shows
the radial profile of $\sigma^2({\rm image}-{\rm psf})$ calculated in
circular annuli of width 0.5\,arcsec. Here the psf has been scaled so
the total intensity is the same for both quasar and psf. As can be
seen, the psf observed with the quasar image matches the quasar close
to its centre better than any other psf. As the core of the psf is
undersampled and the sampling between psf and image has not been
matched (see Section~\ref{sec:nuclear} for further discussion of
this), this result demonstrates the validity of the psf measuring
technique.

\section{The Data Reduction} \label{sec:data}

The data reduction procedure was optimised to search for low surface
brightness extended objects. The same procedure was used for both
image and psf data so no extra differences between measured and actual
psf were introduced. In calculating the flat-field for each mosaic, it
was decided to ignore all pixels within the quadrant containing the
quasar. This ensured that the flat-fielding technique was not biased
to remove or curtail extended emission which could occur if a routine
based on pixel values, such as a $\sigma$-clipping routine, was
used. Outside this quadrant, any areas occupied by bright stars were
also removed from the calculation. The sky background level, assumed
to be spatially constant was also calculated ignoring these areas.

As the images were undersampled in the central regions we decided to
use a sub-pixel shifting routine to centralise the images before they
were median stacked to provide the final composite. Having replaced
bad pixel values, the images were shifted using a bicubic spline
interpolation routine in order to equalise the intensity weighted
centres, and were median stacked. Because the psf quality was found to
vary from day-to-day, the final stacking of psf images was performed
with the same weighting between days as for the quasar images.

Finally, any nearby bright objects in the psf frame were replaced by the
average in an annulus of width 0.5\,arcsec at the distance of the
object from the centre of the psf, around the centre. In the quasar
frame any nearby objects were noted and blanked out of the error frame
so they were not included when measuring $\chi^2$ between image and
model (see Section~\ref{sec:model}).

\section{Determining Simple Morphology} \label{sec:simple}

Because of the often disturbed morphology of quasar hosts it is not
possible to immediately assume a form for the galaxy structure. For
instance if the host is involved in a close merger, modelling it with
a smooth profile will not provide the correct host luminosity. The
extended wings of the psf from the intense nuclear component hide the
host galaxy sufficiently that direct observation cannot easily reveal
even violently disturbed morphologies. Simply subtracting a multiple
of the psf from the centre of the image will reveal some structure,
but a deconvolution routine will reveal more structure. The routine
used was a modified Clean algorithm developed for this problem which
will be described elsewhere \cite{percival_clean}. The results show
that this routine was of sufficient quality to reveal the approximate
symmetry of the host on a scale which includes most of the light
important for modelling the galaxy.

\begin{figure*}
\begin{minipage}{\textwidth}
  \begin{center}
    \resizebox{8.0cm}{8.0cm}{\includegraphics{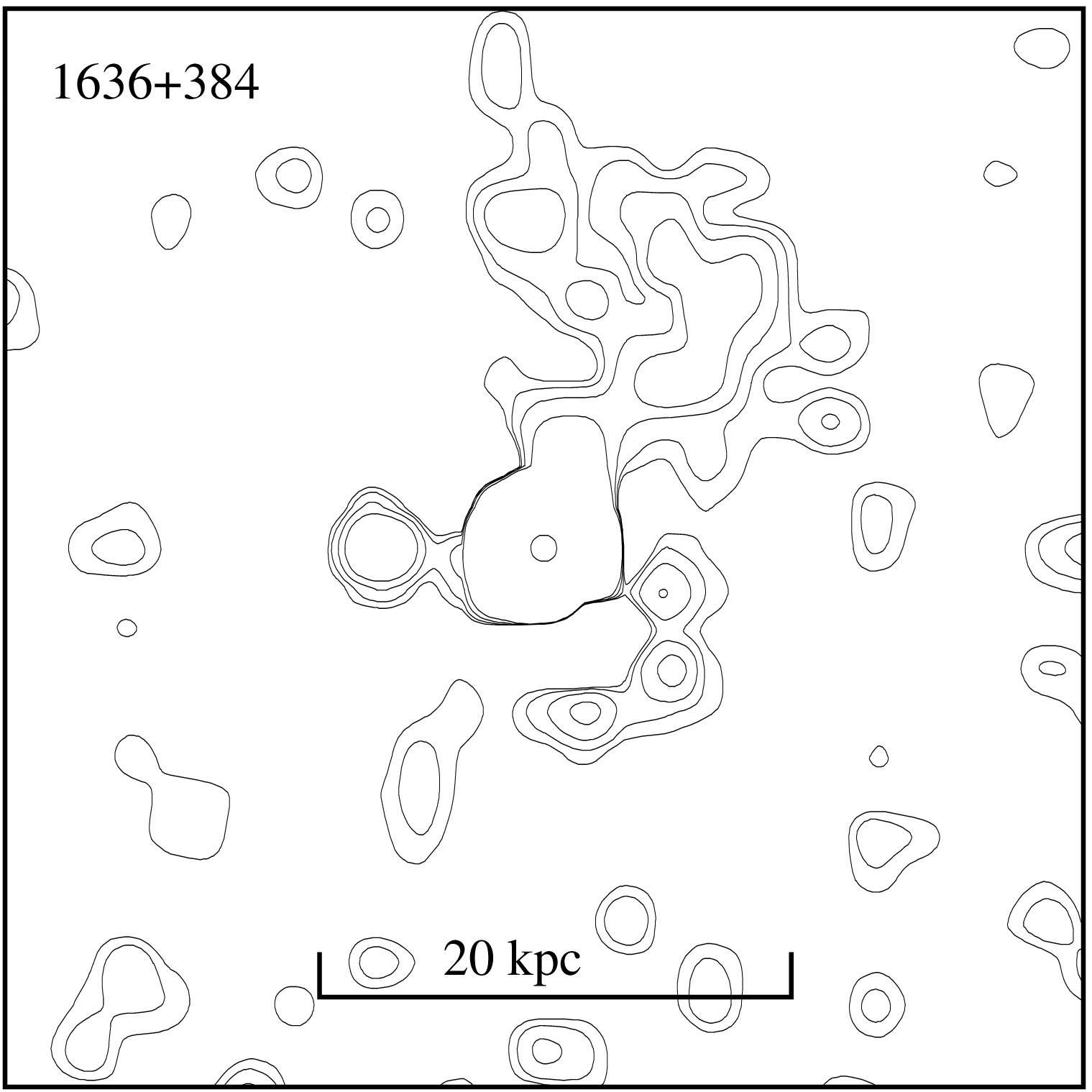}} 
    \hspace{1.0cm}
    \resizebox{8.0cm}{8.0cm}{\includegraphics{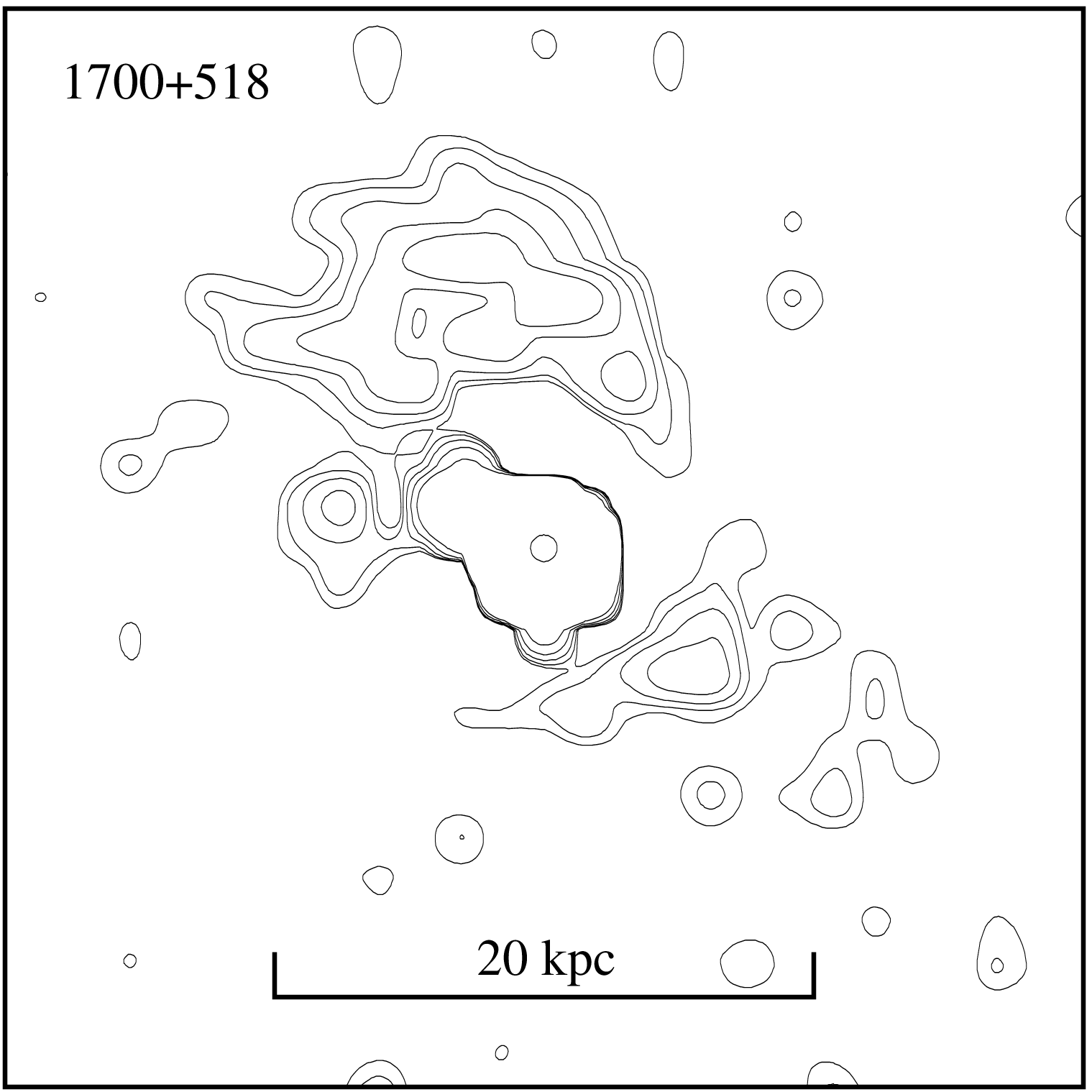}}
  \end{center}
  \centering 

  \caption{The results of the deconvolution of quasars 1636+384 and
  1700+518 revealing hosts with disturbed morphologies indicative of
  close merger events. The deconvolution output, obtained on the pixel
  scale, has been smoothed by convolving with a Gaussian with
  $\sigma=0.5$\,pixels and the residual frame remaining after the
  algorithm has finished had been added back in to preserve
  luminosity. Contours are shown at 0.0125\%, 0.025\%, 0.05\% and 50\%
  of the peak intensity. For quasar 1700+581 a contour is also
  included at 0.00625\% of the peak intensity. The size of the images
  is 7.3$\times$7.3\,arcsec (26$\times$26\,pixels). See
  Section~\ref{sec:results-quasars} for further discussion of the
  morphologies revealed.}
\label{fig:disturbed}
\end{minipage}
\end{figure*}

Examination of the deconvolved images revealed a clear distinction
between disturbed and symmetric systems. Two of the quasars have
morphologies which showed no sign of elliptical symmetry and instead
show signs of recent merger. The deconvolved images of these quasars
are shown in Fig.~\ref{fig:disturbed}. From these images a value for
the non-nuclear luminosity was obtained by summing the residual light
excluding the central pixel. Unfortunately the non-nuclear structure
revealed was not of sufficient quality to be extrapolated into the
central region so the amount of nuclear light which originates in the
host galaxy is unknown. Magnitudes obtained from these deconvolved
images should therefore be treated as approximate. The structure
revealed for these quasars is discussed in
Section~\ref{sec:results-quasars}.  Deconvolving the remaining quasars
revealed host galaxies with approximate elliptical symmetry.

\section{Modelling the Quasar Images} \label{sec:model}

Having determined that the extended structure around a quasar did not
show signs of a disturbed morphology indicative of a close merger and
revealed approximate elliptical symmetry, the luminosity and
morphology of the host galaxy were estimated by fitting model images
to the data. A $\chi^2$ minimisation technique described below was
used to estimate the goodness of fit of the models.

\subsection{Producing a model galaxy} \label{sec:galaxy}

In this Section we describe how the empirical galaxy surface
brightness profile given by Equation~\ref{eq:galprofile} was used to
estimate the contribution from the host to the counts in each
pixel. This had to be done carefully because of the poor sampling of
the images.  The profile given by Equation~\ref{eq:galprofile} has
proved to be an excellent fit to many different types of galaxy
\cite{caon,baggett} and it is assumed that, if the hosts are not
undergoing violent merger, this profile provides a good representation
of the galaxy light.

Before the method is described, it is useful to revise how an image is
obtained from the light emitted by the quasar. Initially, the
continuous distribution of light is altered by the atmosphere and the
optics of the telescope in a way approximately equivalent to
convolution with a continuous point spread function. The resulting
continuous distribution is sampled by the detector which integrates
the light over each pixel. This is equivalent to convolving the light
with a square function of value 1 within a pixel and 0 otherwise, and
sampling the resulting distribution assuming uniform response across
each pixel. The dithering and subsequent stacking of the images will
provide another convolution, although by sub-pixel shifting the images
prior to stacking, the effective smoothing width of this function is
reduced to less than 1\,pixel. The whole process can therefore be
thought of as convolving the true psf, the quasar light and a narrow
smoothing function (of width $\sim$1\,pixel) and sampling the
resulting continuous image on the pixel scale.

Because the psf measurements were obtained using exactly the same
procedure as the quasar images, the measured psf is the result of a
convolution of the true psf with the narrow smoothing function.
Convolution is commutative and associative so this smoothing function
is accounted for in the measured psf and further smoothing of the
model galaxy is not required. For this reason the unconvolved model
galaxy should not be obtained by simply integrating the model profile
over each pixel. Sampling the model galaxy profile onto a grid with
spacing equivalent to the pixel scale and convolving with the psf will
not produce a correct model galaxy because of aliasing.

In order to limit the aliased signal, the procedure adopted was to
extrapolate the psf onto a grid which was finer than the pixel scale
using a sinc function (so no extra high frequency components are
introduced). The surface brightness of the model galaxy was then
calculated at each point on a grid of the same size and was convolved
with the psf on this grid. To provide the final model, this
distribution was subsampled onto the pixel scale. Progressively finer
grids were used until the total counts in the sampled model galaxy
converged, when the majority of the aliased signal is assumed to have
been removed. The algorithm adopted used a fine grid with 4$\times$
the number of points at each successive step, and was stopped when the
average of all the counts differs from that of the previous step by a
factor less than 0.01.

Unless stated otherwise, all model host luminosities and magnitudes
which relate to a 2D profile should be assumed to have been integrated
to infinite radius. For the large radius within which such models were
fitted to the data, this makes only a small difference in the
luminosity. The four parameters of the host galaxy are the geometric
radius of the elliptical annulus which contains half of the integrated
light $R_{1/2}$, the total integrated host luminosity $L_{\rm int}$,
the projected angle on the sky $\alpha$, the axial ratio $a/b$ and the
power law parameter $\beta$. In this paper, the integrated host
luminosity is quoted in counts /analogue data units (adu) detected in
a 1\,sec exposure.

\subsection{The nuclear component} \label{sec:nuclear}

In principle, adding in the nuclear light is simple - the correct
amount is added to the centre of the model galaxy to minimise $\chi^2$
between model and observed images. However, it is important to account
for all of the nuclear light. Differences between measured and true
psf caused by undersampling, seeing variations or effects such as
telescope shake must be accounted for, even though the adopted
observing strategy has limited some of these. In particular, when a
nearby star (as used in Section~\ref{sec:star}) is deconvolved, the
resulting light appears not only in the central pixel, but in the
surrounding pixels as well: if similar components from the nuclear
light are not accounted for in the quasar images, the host
luminosities and morphologies derived will be wrong.

Because of the large peak in both the image and psf, trying to alter
the sampling of the images by extrapolating onto a fine grid and
resampling without inducing unwarranted frequency components causes
`ringing' in the images which is large enough to affect the results of
the modelling. It would be possible to use a different extrapolation
technique, but this risks altering the image and measured psf in
different ways. Instead, a more simple correction to this problem is
adopted: rather than only adding a multiple of the psf to the central
pixel, variable multiples of the psf are also added centred on the
surrounding 8 pixels. In the perfect case where the measured psf is
accurate, while such free parameters make convergence to the minimum
$\chi^2$ value slower, they do not affect the position of the minimum:
the additional components make a negligible contribution to the
model. However, suppose there is a discrepancy between measured and
true psf so that the deconvolved image of the nuclear light consists
of a central spike surrounded by corrective components which decrease
in magnitude with distance away from the spike. Allowing the value of
the pixels close to the core to be free parameters in our model will
correct these discrepancies and any light observed originating away
from the core will be more likely to be from the host galaxy and not
from escaping nuclear light. The opposite is also true, and these
extra components will also correct psf measurement errors which cause
light from the galaxy to be wrongly ascribed to the core (as for
quasar 0956$-$073: see below). Undersampling problems do not affect
the modelled galaxy to the same extent because the galaxy light is
more uniformly distributed and discrepancies are smoothed. In
particular, the total integrated light measured to be from the host
will only be minimally affected: see below.

Quasar 0956$-$073 was modelled using different numbers of these extra
components, and the recovered host parameters are given in
Table~\ref{tab:nuc_param}. As expected, the $\chi^2$ value decreases
with an increasing number of additive psf components showing that the
fit between model and image is being improved. The recovered
parameters for 1 or 9 additive components show moderate differences,
but allowing more components makes no further significant
change. Because the host luminosity increases for this quasar with
increasing numbers of added components, the sum of the extra psf
contributions must be negative which suggests that, for this quasar,
the psf has a slightly broader central profile than the quasar.

For all of the quasars modelled, the total light within the eight
extra psf components was not found to be systematically positive or
negative. If adding 8 extra `psf components' around the core had
always resulted in a total positive (or negative) component being
subtracted from the quasar, this would have suggested that either
these components were removing host light in addition to `leaking'
nuclear light, and that the host profile breaks down in a systematic
way for these pixels, or that our observing strategy had produced a
systematically incorrect psf.

For all quasar host galaxy studies, there is no escaping the
fundamental problem that the galaxy profile has to extrapolated into
the central region from some radius (to separate host and nuclear
light).  By adding in these extra components, all we're doing is
extrapolating from different distances, and arguing that simply
extrapolating only into the central pixel is not necessarily correct
for these data. This is because the measured psf is incorrect for the
(discrete) deconvolution problem we're trying to solve.

\begin{table}
  \centering \begin{tabular}{ccccccc} \hline
  psf   & $R_{1/2}$ & $L_{\rm int}$ & a/b & $\alpha$ & $\beta$ & 
    $\chi^2$ \\
  cmpts & /kpc      & /adu & & &  & \\ \hline
  1  & 10.45 & 312.7 & 0.61 & 1.05 & 0.77 & 2840.3 \\
  9  & 9.76  & 350.1 & 0.64 & 1.04 & 0.92 & 2814.2 \\
  25 & 9.71  & 353.3 & 0.65 & 1.03 & 0.93 & 2799.9 \\
  \hline
  \end{tabular}

  \caption{Recovered host galaxy parameters for quasar 0956$-$073
  modelled using different numbers of additive psf components.}
  \label{tab:nuc_param}
\end{table}

\subsection{The error frame} \label{sec:error}

Determining the fit between model and image requires an estimate of
the relative noise in each pixel, from both intrinsic noise in the
image and differences between measured and true psf. Ideally these
errors should be estimated without recourse to the images but,
unfortunately, this is impractical for these data. Faced with a
similar problem, Taylor \etal\ \shortcite{taylor} estimated the radial
error profile by measuring the error in circular annuli of the image
from which a multiple of the psf had been subtracted centred on the
quasar with matched total luminosity. Using both the image and
measured psf in this way allows the error from psf differences to be
included in the error frame. However, this model assumes that the host
galaxies do not introduce any intrinsic variations in the annuli
within which the variance is calculated. Such variations could result
from either differences between the radial profile of the host
(convolved with the psf) and the psf profile, or significant deviation
from circular host profiles. These effects will be small because the
hosts only contribute a small percentage of the light and deviations
from circular hosts are small.

In order to reduce the number of parameters required to calculate the
error profile and hopefully alleviate any damage caused by calculating
the error frame from the data, Taylor \etal\ \shortcite{taylor} showed
that a function of the form
$\log(\sigma_i)=A\exp^{-0.5(r/S)^{\gamma}}-B$, where A,B,S and
$\gamma$ are four parameters, provides a good fit to the resulting
profile. This profile models both the error in the central regions of
the image and the Poisson background error outside the core. The four
parameters are determined for each quasar by least-square fitting to
the observed error profile. Such a fitting procedure also enables the
error to be determined in the central regions where the gradient is
too steep and there are too few points in each annulus to predict
confidently the error. In general this function fits the observed
error profiles very well, and is used here (without including a
contribution from the host) to estimate the errors in each pixel.

\begin{figure}
  \centering
  \resizebox{\columnwidth}{!}{\includegraphics{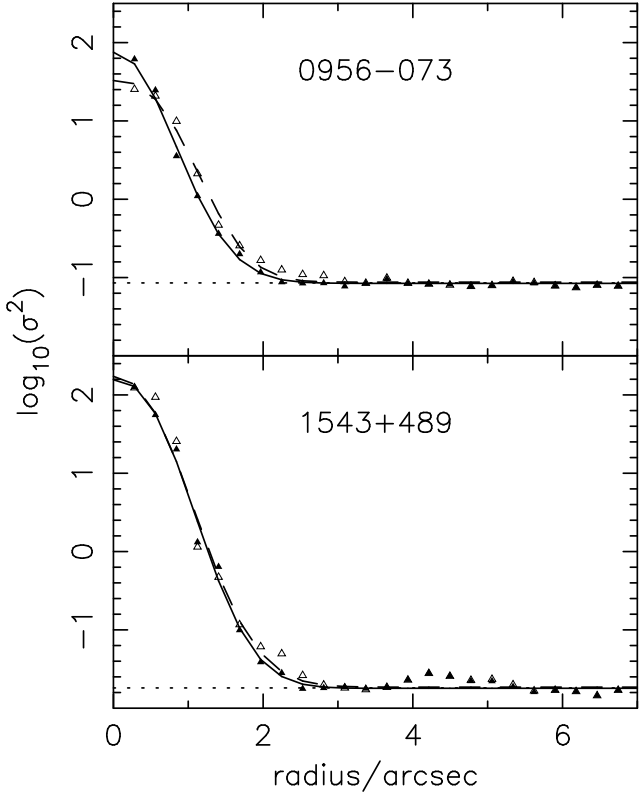}}
  \caption{The calculated radial profile of the variance measured in
  annuli of width 0.3\,arcsec of the difference between the image of
  the quasar and the scaled psf (open triangles). The best-fit model
  for this profile calculated as in Section~\protect\ref{sec:error} is
  also plotted (dashed line). For comparison the radial profile,
  calculated in the same way, for the difference between the image and
  best-fit model galaxy with one nuclear component is plotted (solid
  triangles) and the model of this profile (solid line). This solid
  line shows the radial error profile used to produce the simulated
  data of Section~\ref{sec:mock}. The dotted line shows the Poisson
  noise level of the sky background. Top panel: quasar 0956$-$073,
  bottom panel: quasar 1543+489.}  \label{fig:epgal}
\end{figure}

Fig.~\ref{fig:epgal} shows the observed and fitted error profiles for
quasars 0956$-$073 and 1543+489 with and without including the
best-fit model galaxy in the analysis. The best-fit host around quasar
1543+489 has an axial ratio of 0.89 in contrast to 0.64 for
0956$-$073. The deviation of the host around 0956$-$073 from circular
symmetry explains why the error profile changes when including this
host more than for quasar 1543+489.

\subsection{$\chi^2$ minimisation} \label{sec:minimise}

The algorithm used to find the global minimum in $\chi^2$ was a
multi-dimensional direction set technique based on a method introduced
by Powell in 1964 \cite{nr}. This algorithm requires an initial `start
point' from which it works its way downwards until it finds the
minimum position. Briefly, the algorithm minimises $\chi^2$ by
sequentially adjusting each parameter (i.e. minimising along the axes
of the parameter space), and then it minimises $\chi^2$ on the vector
along which the greatest change was made to $\chi^2$ in the previous
steps. This procedure is repeated until the algorithm
converges. Additionally, for all the quasars it was ascertained that
the algorithm had found the correct minimum and not erroneously
finished early due to a numerical convergence problem by repeatedly
re-running the algorithm starting from the previous best-fit
parameters until the total host luminosity found for successive runs
differed by less than 0.1\,adu. The testing performed for this
algorithm is described in Section~\ref{sec:minima}.

For all the results presented in Section~\ref{sec:results}, the
algorithm was started from an initial position in parameter space
corresponding to a broad, low luminosity galaxy. This was chosen so
the algorithm avoided straying into a region of parameter space where
all of the host light was in the core (i.e small $R_{1/2}$). This is a
relatively flat region for $\chi^2$ in parameter space and it can
therefore take a long time for the algorithm to work its way out of
this region. 

All pixels within a radius of 31\,pixels (8.7\,arcsec) measured from
the centre of the quasar were included in the calculation of
$\chi^2$. For all of the best-fit host galaxies, the difference
between model host luminosity within this area and the integrated
luminosity was negligible, which implies that this area contained all
of the important signal.

\section{Testing the Modelling Procedure} \label{sec:test}

\subsection{Robustness to the error profile}

As a test of the robustness of the best-fit host luminosities to the
determination of the error profile, we have modelled our image of
quasar 0956$-$073 using different error profiles. Quasar 0956$-$073
was chosen for this test because the derived axial ratio of the host
is the lowest of any quasar (although the range of values is quite
small: Section~\ref{sec:axes}). If the galaxy is important in the
error frame calculation, the error profile calculated for this quasar
as in Section~\ref{sec:error} should be the most affected by the fact
that we are ignoring the host (see Fig.~\ref{fig:epgal}). Using an
error profile calculated as in Section~\ref{sec:error} but using the
image only (i.e. not subtracting the psf), the integrated host
brightness was found to drop from 350.1\,adu to 316.6\,adu,
corresponding to a variation of $\sim$0.1\,mag. We have also tried
re-calculating the error frame from the image minus the best-fit model
image (galaxy and nuclear component convolved with the psf), again
using the above formula to fit the error frame. Radial profiles of the
two error frames are shown in Fig.~\ref{fig:epgal}. The best fit model
parameters were used to calculate a new error frame, and we repeated
this process until the best-fit host luminosity converged (subsequent
iterations altered the integrated host luminosity by less than
0.1\,adu). The final best-fit luminosity was found to be 351.5\,adu, a
negligible difference from the original minimum.

\subsection{Finding the minima} \label{sec:minima}

Obvious tests to perform are that there is only one minimum for each
quasar, and that the $\chi^2$ function is well behaved around this
point. Obviously, it is impossible to cover every position in
parameter space to check that $\chi^2$ is well-behaved and that there
are no local minima. However, we have examined the region of parameter
space of interest using a variety of techniques and have found no
potential problems.

The minimisation algorithm is itself designed to cover a large region
of parameter space; the algorithm sequentially searches for the
minimum along a series of vectors (see Section~\ref{sec:minimise} for
details), and considers a large number of diverse values along each
vector. Rerunning the algorithm starting at the best-fit location
previously found also tests any minimum along each axis in parameter
space, as does the calculation of the error bars, described in
Section~\ref{sec:bars}. The shapes of the surfaces around each minimum
are also revealed by this calculation.

\begin{figure}
  \centering \resizebox{\columnwidth}{!}{\includegraphics{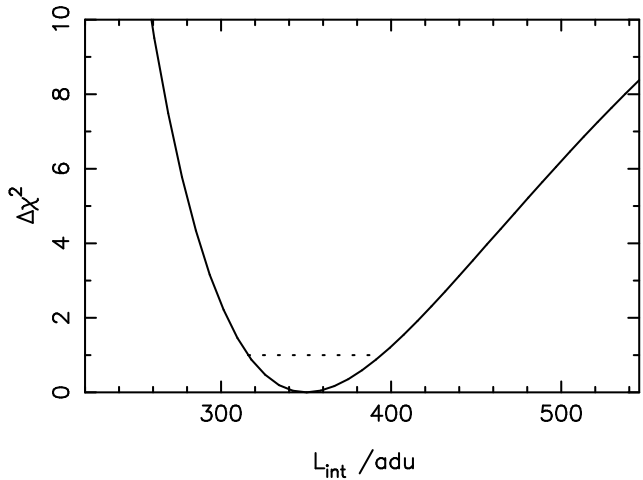}}
  \caption{The variation of $\chi^2$ (normalised to the minimum value)
  versus fixed total host luminosity. For each point on the curve, all
  parameters other than the luminosity have been altered to obtain the
  local minimum in $\chi^2$. The 68.3\% confidence interval, found by
  a separate binary search is also shown (dotted line).}
\label{fig:smooth}
\end{figure}

A test for local minima has been performed for quasar 0956$-$073 over
a larger region of parameter space: the minimisation algorithm was
started at a large number of diverse initial host parameters, and no
significant change in the best-fit parameters was discovered. Quasar
0956$-$073 was chosen for this test because it has average
signal-to-noise of any quasar. Fig.~\ref{fig:smooth} shows a `slice'
through parameter space revealing how smoothly the constrained
$\chi^2$ minimum varies with fixed host luminosity for quasar
0956$-$073. To calculate each point of this curve, all of the
parameters except the host luminosity were varied until the
constrained $\chi^2$ minimum was reached. The remarkable smoothness of
this curve demonstrates both that the global minimum is well
pronounced and the function varies smoothly towards it, and that the
minimisation routine is finding the correct minimum at each point: if
it were not, a more rough surface would be expected, signifying that
the optimum position had not been reached for each host luminosity.

\subsection{Using the $\chi^2$ statistic} \label{sec:chisq}

Use of the $\chi^2$ statistic is dependent on the error in each pixel
being independent of the errors in the other pixels. This is expected
if the errors in the images are dominated by Poisson shot noise. Any
large-scale differences between actual and model host could provide
correlated errors, although these would hopefully have been discovered
by the analysis of Section~\ref{sec:simple}. It is possible that
small-scale discrepancies remain that extend across more than one
pixel. However, the relatively large pixel scale works to our
advantage by reducing the likelihood of this. The central limit
theorem then suggests that the error in each pixel should have
approximately Gaussian distribution.

The minimum $\chi^2$ values are highly dependent on the normalisation
of the error frames, and cannot directly provide tests of the model
fits. The position of these minima are unaffected by the normalisation
of the error frame as they are only dependent on relative variations
between pixels. Examining the reduced $\chi^2$ values at the minima
given in Table~\ref{tab:errors}, we see that the reduced $\chi^2$ is
less than 1 for the majority of the quasars, and deduce that the
procedure outlined in Section~\ref{sec:error} slightly over-estimates
the error in each pixel. This is as expected due to the effect of the
host galaxy. The confidence intervals calculated in
Section~\ref{sec:bars} will therefore be slightly too large, thus
providing a moderately pessimistic error analysis.

As any nearby companions were excluded when measuring $\chi^2$, the
number of pixels used, presented in Table~\ref{tab:errors}, varies
between quasars. For quasar 1214+180, a diffraction spike from a
nearby star which ran close to the quasar was also
excluded. Unfortunately the position of the pixels which were not
modelled is more important than the number of such pixels and, for
this quasar the position of the diffraction spike was such that it
covered a highly important region of pixels. Even though the area
covered was small, the modelling suffered greatly.

\subsection{Calculating error bars on the parameters} \label{sec:bars}

Provided that the galaxy model is a good representation of the true
underlying host galaxy, the errors between the model and image are
uncorrelated between separate pixels, and the procedure in
Section~\ref{sec:error} provides approximately the correct error frame
(see Section~\ref{sec:chisq}), it is possible to calculate error bars
on the true parameters using the $\chi^2$ statistic. The procedure to
do this is to hold the chosen parameter fixed at a certain value, and
minimise the remaining parameters to find the local minimum in
$\chi^2$. The end points of the 68.3\% confidence intervals on the
best-fit parameter are given by the points for which
$\Delta\chi^2=\chi^2-\chi_{\rm min}^2=1$, where $\chi_{\rm min}^2$ is
the minimum value calculated allowing all parameters to vary
\cite{bevington}. A standard binary search has been used to find the
required limits. As well as allowing error bars to be calculated, this
procedure enables the parameter space to be examined and any problems
for each quasar to be spotted.

In order to match the light from the host galaxies, the behaviour of
the integrated host luminosity, $\beta$ and $R_{1/2}$ are coupled
\cite{abraham}. The determination of the error bars is therefore
complicated by the question `what limits, if any should be placed on
the parameters being adjusted to find the constrained minima?'. In
finding the global minima, all of the parameters are effectively
allowed to vary over all space: although bounds are placed on the
parameters, they are not reached (except when modelling the star, see
Section~\ref{sec:star}).  However, at fixed integrated host
luminosity, these limits are often reached because the profiles
required to optimally match the light do not necessarily have to be
those of galaxies. The philosophy adopted is that all the parameters
should be allowed to vary except $\beta$, upon which limits of
$0.25<\beta<6.0$ should be set to provide some adherence to standard
galaxy profiles.

For quasar 0956$-$073, we have examined the required cut through
parameter space for the integrated host luminosity, calculated by
minimising all other parameters to obtain each point. The distribution
of local $\chi^2$ minima are shown in Fig.~\ref{fig:smooth}: the curve
displays simple structure, monotonically decreasing to the global
minimum from both directions so we are justified in using the simple
$\Delta\chi^2=1$ cut-off for the error bars. The resulting 68.3\%
confidence interval for the luminosity is also shown.

The value of $\chi^2$ depends on the error frame used, and it is
expected that the error bars do so as well. The effect of altering the
error frame for quasar 0956$-$073 has been tested by using the error
frame calculated from the image only as in Section~\ref{sec:error}.
Using this error frame, the 68.3\% confidence interval on the host
magnitude changed slightly from $-25.11<M_{K}<-24.87$ to
$-25.10<M_{K}<-24.66$.

\subsection{Fitting to a normal star} \label{sec:star}

On the same frame as quasar 0043+039 we observed a star of similar
signal-to-noise as the quasar. As a test of the fitting procedure we
decided to see if we could fit a `galaxy' to the star. Starting from
an initial position in parameter space corresponding to a broad, low
luminosity galaxy as adopted in all of the modelling, the resulting
best-fit parameters are given in Table~\ref{tab:chisq}. As can be
seen, the fitting procedure rolled down the hill towards a host galaxy
of very low luminosity. At such low total luminosity, the remaining
four galaxy parameters are poorly determined: altering these
parameters results in a very small change in $\chi^2$. Consequently it
is no surprise to find that the best-fit $\beta=6$ value is one of the
limits set in the modelling procedure.

\section{Results of the Analysis}  \label{sec:results}

\begin{table*}
\begin{minipage}{\textwidth}
  \centering \begin{tabular}{cccccccccc} \hline 
  quasar & $R_{1/2}$ & $L_{\rm int}$ & axial ratio & $\alpha$ &
  $\beta$ & nuc/host & $K_{\rm host}$ & $K_{\rm tot}$ &  $K_{\rm psf}$ \\
  & /kpc & /adu & & /radians & & ratio & & & \\ \hline

  0043+039 & 7.41  & 355.0    & 0.96 & 1.44 & 0.60 & 13.9 & 16.04 & 13.54 & 11.90 \\
  0137$-$010 & 5.09  & 407.5  & 0.66 & 2.55 & 2.05 & 13.7 & 15.96 & 13.46 & 10.83 \\
  0244$-$012 & 3.28  & 165.6  & 0.86 & 2.66 & 0.61 & 16.6 & 16.87 & 14.18 & 11.15 \\
  0316$-$346 & 8.56  & 864.8  & 0.76 & 0.07 & 1.25 & 12.2 & 15.15 & 12.73 & 9.62  \\
  0956$-$073 & 9.76  & 350.1  & 0.64 & 1.04 & 0.92 & 21.4 & 16.02 & 13.07 & 9.20  \\
  1214+180 & 3.24  & 259.7    & 0.74 & 1.25 & 1.12 & 15.2 & 16.34 & 13.75 & 8.52  \\
  1216+069 & 19.75 & 379.3    & 0.74 & 2.42 & 2.73 & 21.4 & 15.93 & 12.99 & 11.10 \\ 
  1354+213 & 11.55 & 481.4    & 0.69 & 1.38 & 0.73 & 6.79 & 15.67 & 13.82 & 9.21  \\
  1543+489 & 7.75  & 320.1    & 0.89 & 3.00 & 0.67 & 25.9 & 16.12 & 12.98 & 9.92  \\
  1636+384 & -     & -        & -    & -    & -    & 24.1 & 17.65 & 14.11 & 10.07 \\
  1700+518 & -     & -        & -    & -    & -    & 14.8 & 15.86 & 11.82 & 10.23 \\
  2112+059 & 5.59  & 169.1    & 0.93 & 0.49 & 2.13 & 67.0 & 16.85 & 12.72 & 11.08 \\
  2233+134 & 8.01  & 175.0    & 0.80 & 1.54 & 0.64 & 28.5 & 16.88 & 13.64 & 10.17 \\
  2245+004 & 5.16  & 572.9    & 0.81 & 1.93 & 3.16 & 1.25 & 15.52 & 14.87 & 10.02 \\
  star     & 16.6  & 0.000031 & 0.16 & 2.26 & 6.0  & -    & -    & -    & -    \\ 
  \hline  \end{tabular}

  \caption{Best-fit host galaxy parameters as determined by the 2D
  modelling described in Section~\protect\ref{sec:model}. Also
  included for comparison are the best-fit parameters for a nearby
  star found on the frame of quasar 0043+039 which had similar signal
  to noise as the quasar. For quasars 1636+384 and 1700+518
  deconvolution of the images revealed a highly disturbed morphology
  which extended close to the core of the quasar and model fitting was
  not attempted.  Consequently host galaxy magnitudes presented for
  these quasars are the relatively inaccurate measurements calculated
  from the deconvolved images as described in
  Section~\protect\ref{sec:simple}. Nuclear-to-host ratios are
  calculated in the rest frame of the quasar from the derived absolute
  magnitudes in order that these values are consistent with
  Fig.~\ref{fig:hostvnuc} (see Section~\ref{sec:calc_lum} for
  details). The apparent magnitudes of the quasar, the host component,
  and the star used to give a psf measurement are also presented.}
  \label{tab:chisq}
\end{minipage}
\end{table*}

\begin{table*}
\begin{minipage}{\textwidth}
  \centering \begin{tabular}{ccccccccccc} \hline 

  quasar & $\chi^2$ & number of & reduced & 
    \multicolumn{3}{c}{$L_{\rm int}$ /adu} &
    \multicolumn{3}{c}{$M_K$(host)} & $M_K$(tot) \\
  & & pixels modelled & $\chi^2$ & 
    min & best-fit & max & min & best-fit & max & \\ 
    \hline 

  0043+039 & 2639.79 & 2885 & 0.92 & 287.7  & 355.0    & 454.4  &
    $-$25.56 & $-$25.29 & $-$25.07 & $-$28.22 \\
  0137$-$010 & 2661.25 & 2987 & 0.89 & 333.7  & 407.5    & -      &  
    -     & $-$25.08 & $-$24.87 & $-$28.01 \\
  0244$-$012 & 2985.64 & 2974 & 1.00 & 63.9   & 165.6    & 999.4  & 
    $-$26.85 & $-$24.90 & $-$23.86 & $-$28.01 \\
  0316$-$346 & 2674.08 & 2987 & 0.90 & 766.3  & 864.8    & 1009.7 & 
    $-$25.61 & $-$25.44 & $-$25.31 & $-$28.24 \\
  0956$-$073 & 2814.17 & 2987 & 0.94 & 315.8  & 350.1    & 394.4  & 
    $-$25.11 & $-$24.98 & $-$24.87 & $-$28.36 \\
  1214+180 & 2499.93 & 2509 & 1.00 & 112.7  & 259.7    & -      & 
    -      & $-$24.94 & $-$24.03 & $-$27.96 \\
  1216+069 & 2437.36 & 2625 & 0.93 & 290.2  & 379.3    & 604.0  & 
    $-$25.62 & $-$25.11 & $-$24.82 & $-$28.49 \\
  1354+213 & 2281.14 & 2987 & 0.76 & 462.4  & 481.4    & 502.1  & 
    $-$25.20 & $-$25.16 & $-$25.11 & $-$27.38 \\
  1543+489 & 2567.80 & 2874 & 0.89 & 263.6  & 320.1    & 402.3  & 
    $-$25.56 & $-$25.31 & $-$25.10 & $-$28.89 \\
  2112+059 & 2876.00 & 2945 & 0.98 & 37.7   & 169.1    & -      & 
    -      & $-$24.91 & $-$23.29 & $-$29.50 \\
  2233+134 & 2679.73 & 2941 & 0.91 & 108.7  & 175.0    & 269.2  & 
    $-$24.58 & $-$24.11 & $-$23.59 & $-$27.78 \\
  2245+004 & 2584.98 & 2987 & 0.87 & 438.1  & 572.9    & 941.6  & 
    $-$26.24 & $-$25.70 & $-$25.41 & $-$26.48 \\
  star     & 3055.42 & 2969 & 1.03 & -      & 0.000031 & -      & 
    -      & -      & -     & -      \\ \hline
   \end{tabular}

  \caption{Table showing the end points of the 68.3\% confidence
  intervals calculated as in Section~\protect\ref{sec:bars} for the
  best-fit integrated host luminosities. To further aid comparison
  between model fits for different quasars, the $\chi^2$ values for
  the best fit model are presented with the number of pixels used in
  this calculation. Note that the position of the un-modelled pixels
  is more important than the number of such pixels.}
  \label{tab:errors}
\end{minipage}
\end{table*}

\subsection{Luminosities} \label{sec:calc_lum}

For three of the quasars, analysis of how $\chi^2$ varies within the
parameter space revealed that the best-fit host luminosity was not
well constrained. A host galaxy was determined as being present in
that a lower limit was determined in all cases. However, the maximum
light which could have come from the host was not clear because the
shape of the host was not sufficiently resolved. The morphology of the
best-fit galaxy at large $L_{\rm int}$ could alter to place the
majority of the host light in the central region. This effect could
have been avoided by placing limits on $R_{1/2}$ or, for instance,
using the near-infrared Fundamental Plane \cite{pahre}, although these
upper limits would have been highly dependent on the criteria set.
The host luminosity is ultimately limited by the total light in the
image, and it is expected that the host luminosities for these quasars
do have upper bounds at high values of $L_{\rm int}$, but these high
values would not be of any use in determining the actual host light.

For the remaining nine quasars, the minima were sufficiently
constrained to provide 68.3\% confidence intervals. Comparison of
different confidence intervals provided information on the depth of
the valleys within which each minimum was found and the quality of each
determination.

\begin{figure*}
  \centering \resizebox{10cm}{!}{\includegraphics{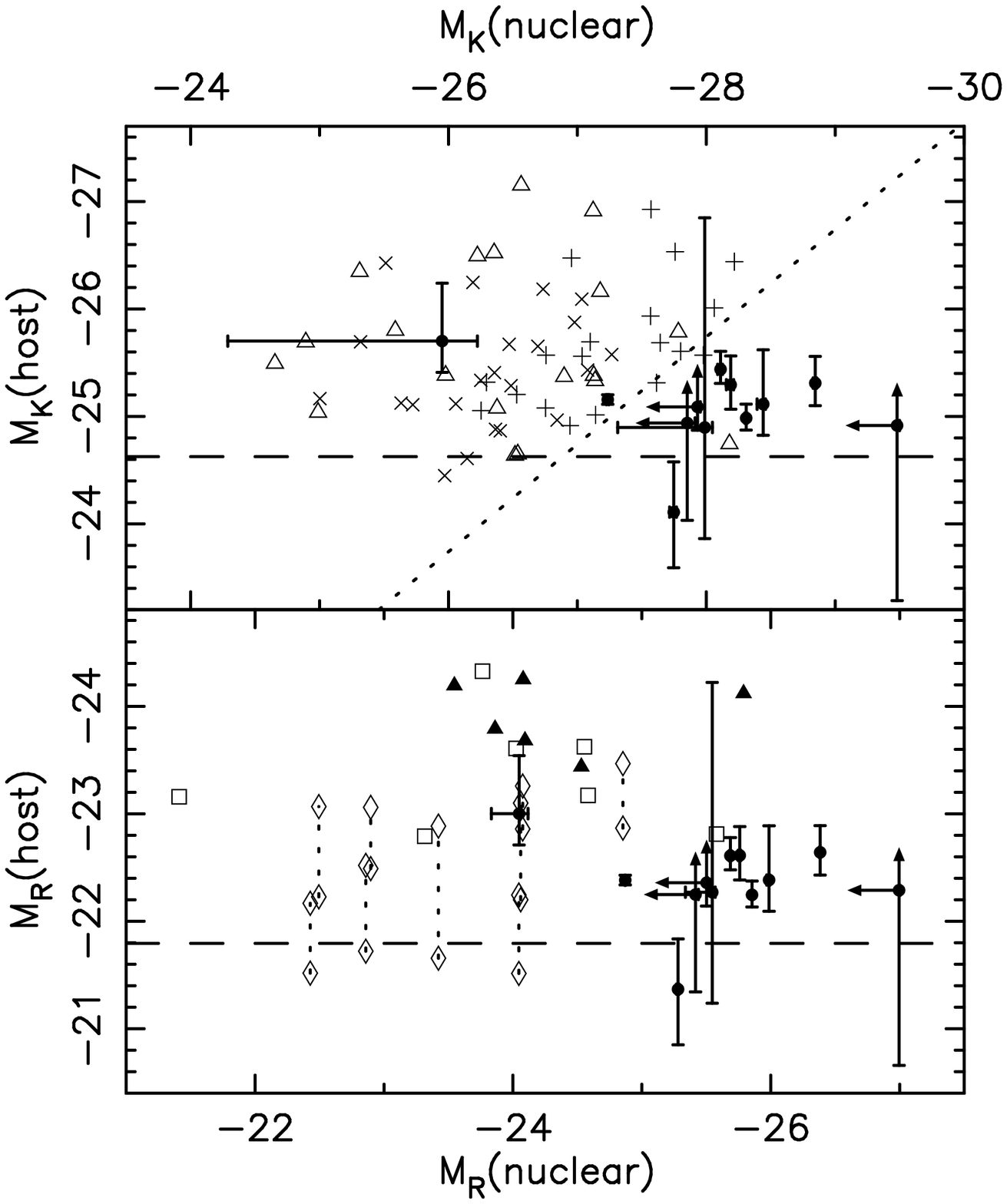}}
  \caption{The top panel shows nuclear vs. integrated host absolute
  $K$-band magnitudes for our sample of quasars (solid circles) with
  error bars calculated as described in Section~\ref{sec:bars}. The
  errors in the measured nuclear component are derived from these and
  consequently the errors are strongly correlated.  Plotted for
  comparison are the calculated host magnitudes for the radio-quiet
  quasars imaged by Taylor \etal\ \protect\shortcite{taylor} (open
  triangles), McLeod \& Rieke \protect\shortcite{mcleod94a} (crosses)
  and McLeod \& Rieke \protect\shortcite{mcleod94b} (plus
  symbols). Details of the conversion of the McLeod \& Rieke data from
  the $H$-band to the $K$-band can be found in
  Section~\ref{sec:calc_lum}. The luminosity of an $L^*$ galaxy,
  $M_K^*=-24.6$ \protect\cite{gardner} is also plotted (dashed line),
  as is the locus of points with a rest-frame $K$-band nuclear-to-host
  ratio of 8 (dotted line). In order to compare with recent HST
  results, in the bottom panel we convert our data into the R-band
  (see Section~\ref{sec:calc_lum} for details). Symbols for our data
  are as for the top panel. The $R$-band best-fit luminosities from
  disk or exponential galaxies \protect\cite{hooper} are also plotted
  (diamonds separated by dotted lines). No attempts were made to
  distinguish the best host morphology in this work. The derived
  $R$-band host luminosities of McLure \etal\
  \protect\shortcite{mclure} are also shown (radio-quiet sample: open
  squares, radio-loud sample: solid triangles). The luminosity of an
  $L^*$ galaxy, $M_R^*=-21.8$ \protect\cite{lin} is plotted for
  comparison (dashed line). }
\label{fig:hostvnuc}
\end{figure*}

Host and nuclear luminosities for our quasars are compared with the
results of other studies in Fig.~\ref{fig:hostvnuc}. In order to
compare with the $H$-band host galaxy studies undertaken by McLeod \&
Rieke (1994a;b), we convert their total (nuclear + host) and host
luminosities to the $K$-band by applying a single conversion factor to
the apparent magnitudes. This then sets the relative normalisation of
the $K$-band and $H$-band samples; conversion to absolute magnitudes
is subsequently undertaken in exactly the same way for all of the
infra-red samples.

In a study of the energy distribution of the PG quasars (from which
McLeod \& Rieke chose their samples), Neugebauer \etal\
\shortcite{neugebauer87} found $\langle H-K\rangle=0.90$ for the
sample of Mcleod \& Rieke \shortcite{mcleod94a} and $\langle
H-K\rangle=0.98$ for McLeod \& Rieke \shortcite{mcleod94b}. In the
upper panel of Fig.~\ref{fig:hostvnuc}, we adopt these values to
convert the total luminosities of the McLeod \& Rieke quasars into the
$K$-band.

The light from the galaxy component is assumed to be dominated by an
evolved stellar population, the colour of which reddens with increasing
redshift. For nearby galaxies, $H-K\sim0.25$, which was used by McLeod
\& Rieke \shortcite{mcleod95a} to convert galaxy absolute
magnitudes. However, the apparent $H-K$ is dependent on redshift and,
at the redshifts of the quasars imaged by McLeod \& Rieke (1994a;b),
$H-K\sim0.6$ is expected for an evolved stellar population
\cite{lilly}. This was adopted to convert the McLeod \& Rieke galaxy
luminosities into the $K$-band.

We have also checked the calibration of the McLeod \& Rieke samples
and our sample of modelled quasars (with 6 overlapping objects)
against the data of Neugebauer \etal\ \shortcite{neugebauer87}. The
average total quasar luminosity for the subsamples are in good
agreement, although individual values vary by up to $0.7$\,mag,
presumably due to intrinsic quasar variability.

One quasar ($1354+213$) was imaged by McLeod \& Rieke
\shortcite{mcleod94b}, Neugebauer \etal\ \shortcite{neugebauer87} and
in our study. Neugebauer \etal\ \shortcite{neugebauer87} derived
$H-K=1.0$ for this object, which is higher than $H-K=0.3$ derived
by combining the McLeod \& Rieke $H$-band and our $K$-band
observation. However, the McLeod \& Rieke and our observations were
undertaken at different epochs, and the luminosity is not expected to
remain constant.

The study of Taylor \etal\ \shortcite{taylor} was performed in the
$K$-band, and the apparent $K$-band magnitudes of host and nuclear
components were taken directly from this work. The data from the
different infra-red samples were then converted to absolute
magnitudes, applying the $K$-correction of Glazebrook \etal\
\shortcite{glazebrook95} for the host galaxy and assuming the nuclear
component follows a standard power law spectrum $f(\nu)=\nu^{-0.5}$.

Using the error bars calculated in Section~\ref{sec:bars} to weight
the data, the average integrated host galaxy magnitude for our quasars
was found to be $\langle M_K \rangle =-25.15\pm0.04$. For comparison,
when converted for cosmology exactly as our data, the sample of Taylor
\etal\ \shortcite{taylor} gives $\langle M_K \rangle =-25.68$, McLeod
\& Rieke \shortcite{mcleod94a} $\langle M_K \rangle =-25.42$ and
McLeod \& Rieke \shortcite{mcleod94b} $\langle M_K \rangle =-25.68$.

Recent determinations of the $K$-band luminosity of an $L^*$ galaxy
\cite{gardner} have resulted in $M_K^*=-24.6$, compared to previous
determinations of $M_K^*=-24.3$ \cite{glazebrook95} and $M_K^*=-25.1$
\cite{mobasher}. The Gardner \etal\ \shortcite{gardner} value is
plotted in the top panel of Fig.~\ref{fig:hostvnuc}. This shows that
the average luminosity of our hosts is $\sim1.6$ times that of an
$L^*$ galaxy. Note that for all three values, the derived average
luminosity is $1-2$ times that of an $L^*$ galaxy, and the conclusions
of Section~\ref{sec:discuss:lum} are not affected by this choice.

We compare our sample to recent HST $R$-band results in the lower
panel of Fig.~\ref{fig:hostvnuc} assuming an apparent $R-K=2.5$ for
the total light from our quasars based on the average value for the 6
quasars which overlap our sample and the sample of Neugebauer \etal\
\shortcite{neugebauer87}. The $R-K$ colour of an evolved stellar
population, assumed to dominate the host galaxies, is dependent on the
redshift of the source and, for the redshifts of our sample
($z\sim0.35$), is expected to be $\sim3.5$ \cite{dunlop89}. All the
data (including our data after conversion to apparent $R$-band
magnitudes) presented in the bottom panel of Fig.~\ref{fig:hostvnuc}
were adjusted for cosmology assuming that the nuclear component has a
spectrum of the form $f(\nu)\propto\nu^{-0.5}$, and the galaxy
component has $f(\nu)\propto\nu^{-1.5}$.

\subsection{Morphologies}  \label{sec:morph}

Morphologies are parametrised by the best-fit value of $\beta$:
$\beta=1$ values correspond to disk-like, and $\beta=4$ to spheroidal
profiles. The technique described in Section~\ref{sec:bars} has been
used to reveal how well the $\beta$ parameter is constrained by the
modelling. The result of this analysis is presented in
Table~\ref{tab:beta}. As can be seen, the $\beta$ parameter is well
constrained for fewer quasars than the luminosity and $\chi^2$ error
bars reveal a highly skewed distribution for the expected true value
given the best-fit $\beta$ value. In order to correctly determine the
differential probability between disk and spheroidal profiles, we need
to know the relative dispersion of $\beta$ for each morphological
type. However, examining the best-fit parameters, the error bars on
$\beta$, and the shape of $\chi^2$ surface, on which we have
information from the binary search to find the error bars, we can
infer the best fit morphology for some of the quasars. The suggestion
from this is that luminous radio-quiet quasars can exist in hosts
dominated by either disk-like or spheroidal components. A histogram of
these data is plotted in Fig.~\ref{fig:mock_beta}, where the
distribution is compared to that recovered from simulated data with
exact $\beta=1$ or $\beta=4$ profiles.

\begin{table}
  \centering \begin{tabular}{ccccc} \hline  

  quasar & min $\beta$ & $\beta$ & max $\beta$ & morphology? \\ \hline 
  0043+039   & 0.42    & 0.60 & 0.86   & disk      \\
  0137$-$010 & 1.52    & 2.05 & $>$6.0 & spheroid  \\
  0244$-$012 & $<$0.25 & 0.61 & 1.62   & disk      \\
  0316$-$346 & 1.02    & 1.25 & 3.62   & ?         \\
  0956$-$073 & 0.71    & 0.92 & 1.21   & disk      \\
  1214+180   & $<$0.25 & 1.12 & $>$6.0 & ?         \\
  1216+069   & 1.44    & 2.73 & $>$6.0 & spheroid  \\
  1354+213   & 0.65    & 0.73 & 0.82   & disk      \\
  1543+489   & 0.49    & 0.67 & 0.90   & disk      \\
  2112+059   & $<$0.25 & 2.13 & $>$6.0 & ?         \\
  2233+134   & $<$0.25 & 0.64 & 1.21   & disk      \\
  2245+004   & 2.17    & 3.16 & 5.55   & spheroid  \\
  \hline  \end{tabular}

  \caption{Table showing 68.3\% confidence intervals for the best-fit
  host $\beta$ parameters for the 12 quasars modelled using the 2D
  $\chi^2$ minimising technique (Section~\protect\ref{sec:model}).
  Error bars were calculated as described in
  Section~\protect\ref{sec:bars} with $\beta$ constrained to lie in
  the range $0.25<\beta<6.0$. The morphology of the dominant
  contribution to the host galaxy is also presented, based on the best
  fit $\beta$ parameter and associated confidence interval.}
  \label{tab:beta}
\end{table}

\subsection{Axial ratios and angles} \label{sec:axes}

\begin{figure}
  \centering 
  \resizebox{5.0cm}{5.0cm}{\includegraphics{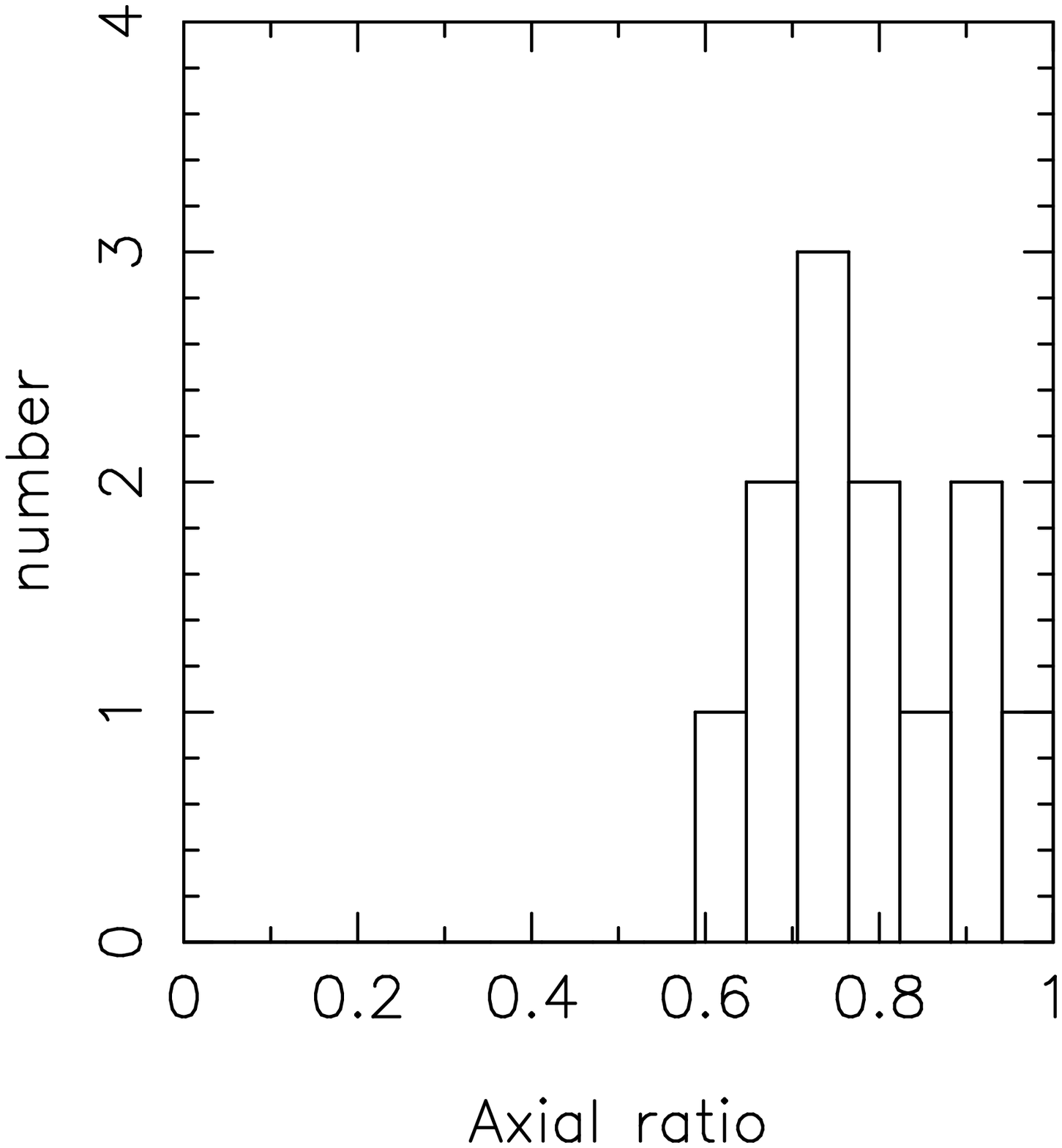}}
  \vspace{5mm}\\
  \resizebox{5.0cm}{5.0cm}{\includegraphics{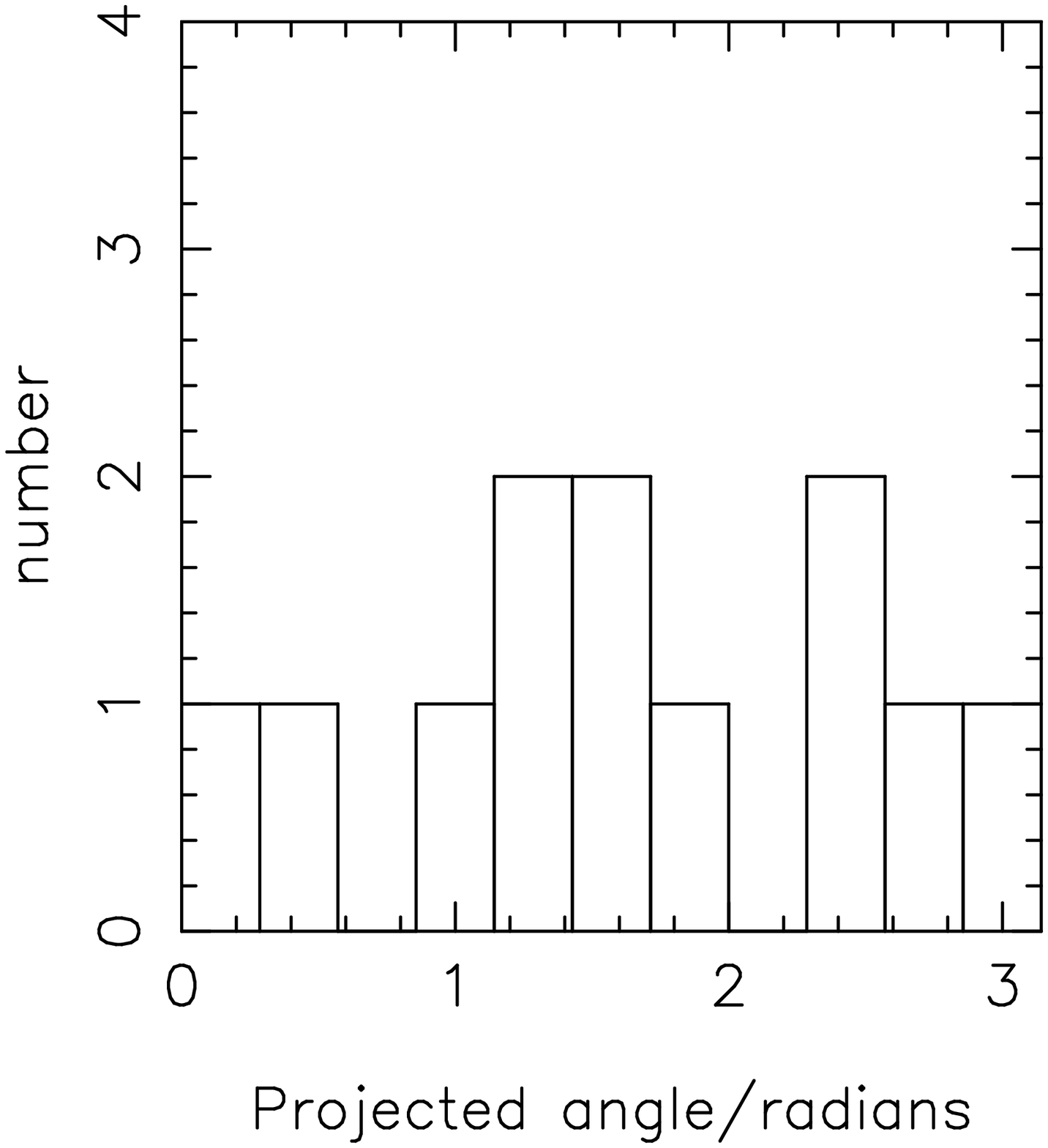}}
  \caption{Histograms showing the distribution of axial ratios and
  projected angles of the host galaxies. These parameters were well
  constrained for all of the 12 quasars modelled, and are therefore
  plotted from all of the minima found. Axial ratios are tightly
  constrained with with $a/b>0.64$ for all hosts, and $\langle
  a/b\rangle=0.79\pm0.03$. The distribution of projected angles is
  approximately uniformly distributed.}
\label{fig:histograms}
\end{figure}

Analysis of the parameter space reveals that the axial ratio and
projected angle of each host are better constrained than the other
parameters. Fig.~\ref{fig:histograms} shows histograms of these
parameters for all quasars modelled. The distribution of axial ratios
is small with $\langle a/b\rangle=0.79\pm0.03$. This is in agreement
with those found by McLure \etal\ \shortcite{mclure}, but higher than
found by Hooper, Impey \& Foltz \shortcite{hooper}. The projected
angles are uniformly distributed as expected.

\subsection{Highlighted results for selected quasars} 
    \label{sec:results-quasars}

\subsubsection{Quasar 0043+039}
The broad-absorption-line (BAL) quasar PG0043+039 has been subject to
2 previous studies to determine host galaxy properties. It was
observed in the $i$ band by Veron-Cetty \& Woltjer
\shortcite{veron-cetty} who determined $M_i=-23.9$ if the host is a
disk like ($\beta=1$) galaxy, or $M_i=-24.7$ for a spheroidal
($\beta=4$) galaxy. This quasar was also observed using the wide-field
camera on HST by Boyce \etal\ \shortcite{boyce}, who used a cross
correlation technique to determine that the host was slightly better
fit by a disk galaxy with $M_V=-21.6$. We also find that the dominant
morphology is disk-like and calculate $M_K=-25.29$. The old burst
model of Bruzual \& Charlot \shortcite{bruzual} predicts $V-K=3.3$
which is consistent with the derived $V-K=3.7$.

\subsubsection{Quasar 0316$-$346}
This quasar was previously observed using the wide field camera on HST
\cite{bahcall} and the host was found to reveal evidence of a merger,
in particular tidal tails extending $\sim20$\,kpc west of the
quasar. Bahcall \etal\ \shortcite{bahcall} also provide a 2D fit to
the host properties and find that the best-fit host is disk galaxy
with $M_{V}=-22.3$. We also calculate a best-fit disk galaxy and find
$M_K=-25.44$, giving $V-K=3.1$, again consistent with the old burst
model of Bruzual \& Charlot \shortcite{bruzual} which predicts
$V-K=3.3$.

\subsubsection{Quasar 1214+180}
There have been no previous attempts to determine the morphology of
the host galaxy around this quasar possibly due to the nearby star
which was utilised in this work to obtain an accurate
psf. Unfortunately in our images, a diffraction spike from this star
passed close to the quasar reducing the area that could be used to
calculate $\chi^2$. Although the modelling converged to give basic
galaxy parameters, further analysis of the parameter space revealed
that this minimum was not well constrained.

\subsubsection{Quasar 1216+069}
Our analysis of this quasar benefited because the images were obtained
using the tip-tilt system and there is a nearby bright star which was
placed on the same frame as the quasar and used to obtain a psf
measurement. Previously, `nebulosity' has been observed around this
quasar \cite{hutchings84}, and a more detailed HST study found a
best-fit spheroidal ($\beta=4$) galaxy with $M_{V}=-22.3$
\cite{boyce}. We also find that the most likely host is a large
spheroidal galaxy and obtain $M_K=-25.1$, giving $V-K=2.8$.

\subsubsection{Quasar 1354+213}
Using a psf subtraction technique, McLeod \& Rieke
\shortcite{mcleod94b} found a residual host galaxy with $M_K=-25.6$
when converted to our cosmology using the $K$-correction from
Glazebrook \etal\ \shortcite{glazebrook95} and the apparent colour
correction $H-K=0.6$ (see Section~\ref{sec:calc_lum}). Our best fit
host luminosity was $M_K=-25.2$. Analysis shows that the luminosity
and the $\beta$ parameter are both tightly constrained by the
modelling and the best-fit $\beta=0.73$ suggests that the host is
dominated by a disk component. The rest-frame nuclear-to-host ratio
for this quasar is only $6.8$ (the apparent nuclear-to-host ratio is
$4.6$), which explains why the derived parameters have small error
bars.

\subsubsection{Quasar 1636+384}
We are not aware of any previous attempts to determine the luminosity
and morphology for the host galaxy of quasar 1636+384. Preliminary
deconvolution of the light revealed that the excess, non-central light
displayed a morphology greatly disturbed from elliptical symmetry (as
shown in Fig.~\ref{fig:disturbed}). The structure includes an excess
of light to the NW of the core which is interpreted as a merging
component as well as light around the central core which probably
originates from the host. From this image it was unclear how to
distinguish between the host and the interacting companion, so the
luminosity of the host was estimated by summing pixel values excluding
the central pixel. This provided an approximate $K$-band absolute
magnitude of $M_K=-23.5$.

\subsubsection{Quasar 1700+518}
Quasar 1700+518 is a bright BAL quasar of low redshift
($z=0.29$). Such low redshift BAL objects are rare and hard to
discover since the broad absorption lines are in the UV and
consequently quasar 1700+518 has received much interest: specific
studies of this quasar have been undertaken in many different
wavebands \cite{h92_1700,stockton,hines}. Because of the low redshift
and the brightness of the quasar, 1700+518 has also been included in
many samples of quasars imaged to obtain details of their host
galaxies \cite{neugebauer,mcleod94b}, although these have only
provided upper limits for the host magnitude. More recent imaging
studies have shown that the morphology of the underlying structure
consists of a disturbed host predominantly to the SW of the core
\cite{stockton} and a close interacting companion to the NE
\cite{h92_1700} which is most likely a ring galaxy \cite{hines}.
Deconvolution of the light from this quasar, as shown in
Fig.~\ref{fig:disturbed}, confirms this picture of the structure. With
the disturbed morphology it is difficult to know how to split the
light in the central pixels into nuclear and host components. As for
quasar 1636+384 the host luminosity was estimated by summing the
counts in the pixels surrounding the central one (ignoring those from
the NE companion). There will be errors caused by leakage of light
from the nuclear component and from the contribution of the host to
the central pixel. An approximate $K$-band absolute magnitude of
$-24.9$ was obtained for the host galaxy and $-24.4$ for the NE
companion galaxy.

\subsubsection{Quasar 2233+134}
Both Smith \etal\ \shortcite{smith} and Veron-Cetty \& Woltjer
\shortcite{veron-cetty} included this quasar in their samples, but
both failed to resolve the host galaxy beyond obtaining upper limits
for the luminosity. Hutchings \& Neff \shortcite{hutchings92} did
resolve the host galaxy and found the host to be best-fit by a
$\beta=4$ model, although they did not resolve further information
about the galaxy. However, we find that the most probable host has an
disk profile and calculate $M_K=-24.1$, the lowest luminosity host
modelled. If we constrain the host to have an elliptical profile, the
best fit luminosity becomes $M_K=-25.9$, although the half light
radius is very small for this model ($R_{1/2}=1.5$\,kpc) which places
it a long way from the $K$-band Fundamental Plane of Pahre, Djorgovski
\& de Carvalho \shortcite{pahre}. If the host parameters are
constrained to lie on this plane, then rerunning the modelling gives a
best fit host with $M_K=-24.6$. Neither of these changes would be
sufficient to alter our conclusions.

\section{Simulated Data I - Single component galaxies}  \label{sec:mock}

Trying to recover known host parameters from Monte-Carlo simulations
of the actual data enables the distribution of recovered parameters
given the true values to be determined. Note that the error bars
calculated using the $\chi^2$ statistic are instead determined from
the distribution of possible true values given the data. These two
distributions are not necessarily equal. We need to determine the
distribution of recovered values in order to answer questions such as
`Are our results biased towards low $\beta$ values?'.

In view of the distribution of recovered $\beta$ values, it was
decided to simulate data to match the images of quasars 0956$-$073 and
1543+489. These quasars span the distribution of signal-to-noise of
all the images, and 2D model fitting revealed evidence for disk
dominated hosts in both cases. Verification of this result is
interesting as recent work has suggested that the hosts of luminous
quasars should be dominated by the spheroidal component (see
Section~\ref{sec:discuss_morph}).

\subsection{Creating the mock data}  \label{sec:mock_noise}

\begin{table*}
\begin{minipage}{\textwidth}
  \centering \begin{tabular}{ccccccccc} \hline  
  quasar & \multicolumn{4}{c}{$\beta$} & 
    \multicolumn{4}{c}{$L_{\rm int}$ /adu} \\
  & true & min & average & max & true & min & average & max \\ \hline

  0956$-$073 & 1.0 & 0.74 & 1.14 & 2.28 & 300.0 & 262.1 & 323.7 & 510.0 \\
  0956$-$073 & 4.0 & 1.93 & 3.77 & 7.42 & 300.0 & 213.6 & 288.3 & 463.1 \\
  1543+489 & 1.0 & 0.79 & 1.08 & 1.69 & 300.0 & 260.1 & 324.7 & 512.2 \\
  1543+489 & 4.0 & 2.13 & 4.04 & 7.25 & 300.0 & 209.6 & 302.2 & 469.8 \\
  \hline 
  \end{tabular}

  \caption{Table showing the mean and 68.3\% confidence intervals for
  the recovered $\beta$ and integrated host luminosity from the
  simulated data. 100 simulations were performed for each morphology
  for each quasar.}  \label{tab:mock_stat}
\end{minipage}
\end{table*}

\begin{figure*}
\begin{minipage}{\textwidth}
  \centering \resizebox{10cm}{!}{\includegraphics{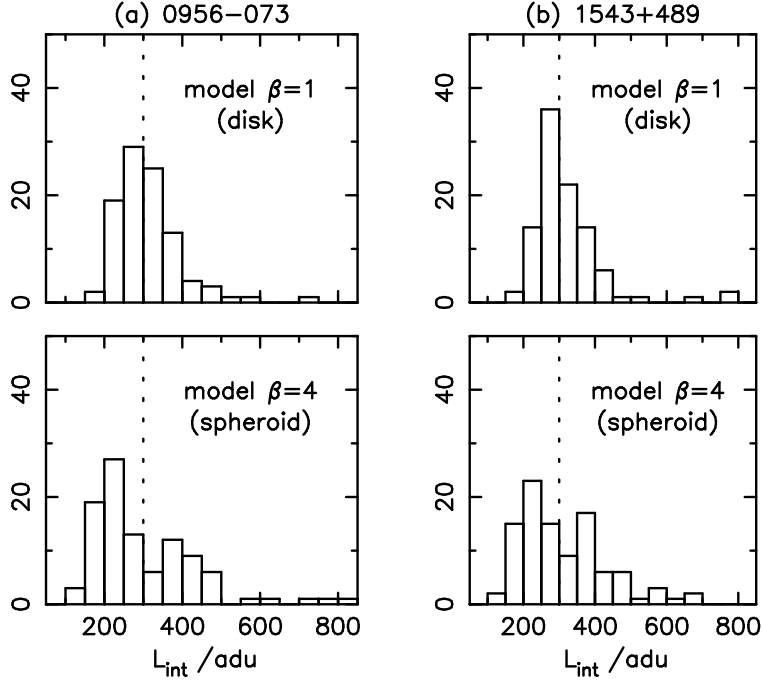}}
  \caption{The distribution of recovered luminosities from Monte-Carlo
  simulations of 100 $\beta=1$ images and 100 $\beta=4$ images. The
  $y$-axis gives the number of recovered values within each luminosity
  bin. Noise has been added to match observations of quasar (a)
  0956$-$073 and (b) 1543+489, including a contribution from the error
  in the measured psf as described in
  Section~\protect\ref{sec:mock_noise}.  The luminosity of each
  simulated galaxy, marked by the dotted line, was set at 300\,adu.}
  \label{fig:mock_lum}
\end{minipage}
\end{figure*}

Simulated galaxies were created using the procedure outlined in
Section~\ref{sec:galaxy} and a single $\delta$ function added to the
centre of each to create a `perfect unconvolved model'. The height of
the $\delta$ function was chosen to match the total signal of the
original images. These models were then convolved with the psf
measured to match the quasar.

Gaussian noise was added with a radially dependent variance as given
by the error profile calculated in Section~\ref{sec:error} including
the best-fit host galaxy in the calculation. The error profiles used
for quasars 0956$-$073 and 1543+489 are given by the solid lines in
Fig.~\ref{fig:epgal}. Differences between measured and true psf were
included in this analysis, and are therefore included in the noise
levels added to the simulated data. This noise model assumes that the
errors in different pixels are independent (see
Section~\ref{sec:chisq}).

We have simulated 100 images with exact disk hosts, and 100 images
with exact spheroidal hosts for each of the two quasars chosen. The
true integrated host luminosity was set at 300\,adu for simulated data
of both quasars. This conservative value is below the best-fit value
obtained from the data for both quasars, providing a stringent test of
the modelling. This is particularly true for a $\beta=4$ host:
constraining $\beta=4$ when modelling the observed image would have
resulted in a best-fit $L_{\rm int}\gg300$\,adu. The simulated images
were analysed using exactly the same 2D modelling procedure described
above for the observed data.  The range of recovered parameters is
analysed below.

\subsection{Results from the simulated data: luminosities}

Recovered luminosities, presented in Fig.~\ref{fig:mock_lum} reveal a
skewed distribution, particularly for hosts with exact spheroidal
profiles where the recovered luminosity is biased towards a low
value. This is consistent with the morphology being skewed towards a
low $\beta$ value (see next Section): if $\beta$ is decreased, the
luminosity also has to decrease to keep the counts in the outer pixels
(those most important for fitting the host) the same. The counts in
the centre of the galaxy are less important because of the additional
nuclear component which is adjusted to match the data.

The mean and variance in the recovered luminosities are presented in
Table~\ref{tab:mock_stat}. Although the error bars reveal the extent
of the skewed distribution, the mean is within 10\% of the true value
for each quasar and morphology.

\subsection{Results from the simulated data: morphologies}

\begin{figure*}
\begin{minipage}{\textwidth}
  \centering \resizebox{15cm}{!}{\includegraphics{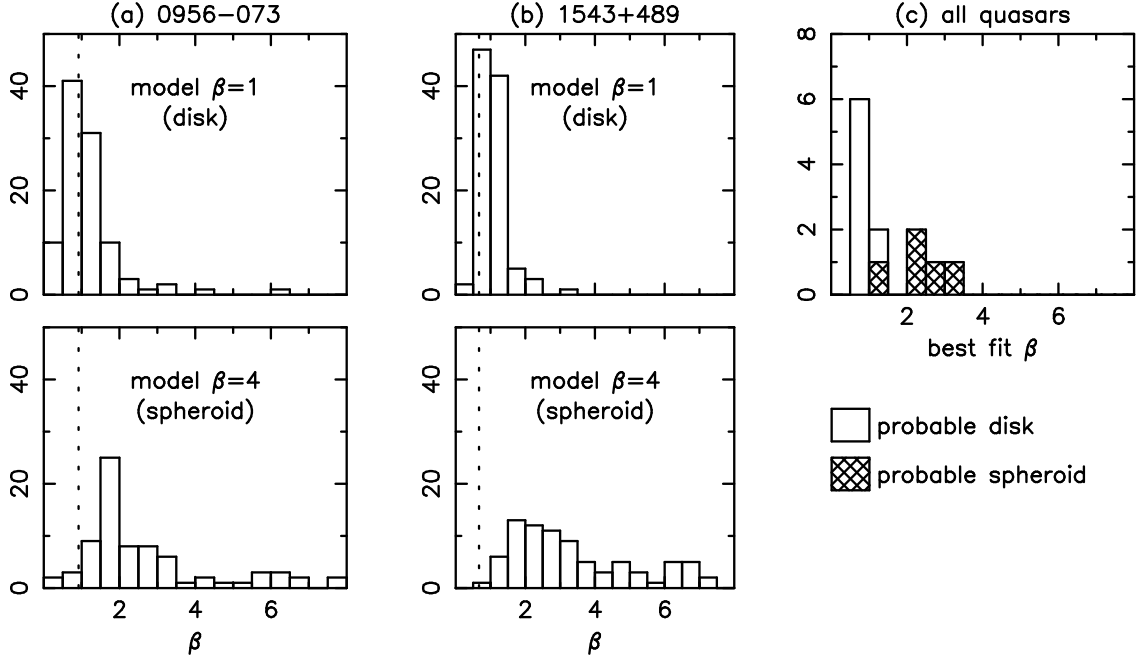}}

  \caption{Evidence that the hosts of luminous radio-quiet quasars are
  not exclusively dominated by spheroidal components. The distribution
  of $\beta$ values recovered from 2-D modelling of 100 simulated
  images created with hosts using exact disk or spheroidal profiles is
  presented. Noise has been added to these images to match
  observations of (a) quasar 0956$-$073 and (b) quasar 1543+489,
  including a contribution from the error in the measured psf as
  described in Section~\protect\ref{sec:mock_noise}. The dotted line
  shows the best-fit $\beta$ value recovered from the images of these
  quasars. (c) For comparison the distribution of best-fit $\beta$
  values obtained from all of our $K$-band images is also plotted. The
  probable morphology of the host was determined from the $\chi^2$
  error bars derived for the true value given the data.}
  \label{fig:mock_beta}
\end{minipage}
\end{figure*}

The skewed distribution observed in the error bars on the true host
$\beta$ value is mirrored by the distribution of $\beta$ values
recovered using the standard modelling procedure described in
Section~\ref{sec:model}. Fig.~\ref{fig:mock_beta} shows the relative
distribution of $\beta$ values retrieved from the simulated images.
Limits of $0.25<\beta<8$ were placed on fitted $\beta$ values. For
quasar 0956$-$073, 16 of the simulated images created with exact
spheroidal hosts, had recovered $\beta>8$. For quasar 1543+489, this
number was 14: these values are not included in
Fig.~\ref{fig:mock_beta}. The distribution was used to calculate the
mean and standard deviation given in Table~\ref{tab:mock_stat},
assuming all fits with $\beta>8$ actually had $\beta=8$.

If the host were a spheroidal galaxy with $\beta=4$, the probability
of recovering a best-fit value of $\beta<1$ is $\sim0.03$ for
0956$-$073 and $\sim0.01$ for 1543+489: the best-fit values from the
images were $\beta=0.92$ and $\beta=0.67$ respectively. The evidence
for the existence of hosts dominated by a disk component therefore
appears to be strong. In Fig.~\ref{fig:mock_beta}, the distribution of
retrieved $\beta$ values for the 12 quasars modelled is also
shown. This distribution is inconsistent with the hypothesis that all
the hosts are dominated by spheroidal components on the scales probed
by these measurements. The histogram is divided to show the probable
distribution of morphologies given the options $\beta=1$ or
$\beta=4$. As can be seen, the modelling suggests that approximately
half of the hosts are dominated by disk components.

\section{Simulated Data II - Two component galaxies} \label{sec:mock2}

In order to constrain the potential importance of spheroidal cores in
the galaxies found to be dominated by disk-like profiles, we have
analysed synthetic quasars created with two host galaxy
components. Using the Fundamental-Plane (FP) relation between
$R_{1/2}$ and $L_{\rm int}$ found in the K-band by Pahre \etal\
\shortcite{pahre}, we have added extra spheroidal ($\beta=4$)
components to the recovered best-fit host galaxy of quasar
1543+489. Note that this best fit host had $\beta=0.67$. We have tried
the same analysis using $\beta=1$ and found no change in the effects
produced by the spheroidal core. After adding in the nuclear component
and noise as described in Section~\ref{sec:mock}, we have recovered
the best-fit host galaxy parameters using our single component
modelling. Spheroidal components were added with a variety of
different luminosities, and five different realisations of the
additional noise component were added to each. The resulting average
recovered $L_{\rm int}$ \& $\beta$ are given in Table~\ref{tab:2cmpt}.

\begin{table}
  \centering
  \begin{tabular}{cccccc} \hline
  \multicolumn{4}{c}{$L_{\rm int}$ /adu} & $R_{1/2}$ & $\beta$ \\
  spheroidal & total & recovered & diff & /kpc & \\ \hline
  0.0	& 320.1 & 291.2  & -28.9 & 8.26 & 0.61 	\\
  40.0	& 360.1 & 306.5  & -53.6 & 8.19 & 0.63 	\\
  80.0	& 400.1 & 337.9  & -62.2 & 8.02 & 0.69	\\	
  120.0	& 440.1 & 382.7  & -57.4 & 7.71 & 0.81	\\
  160.0	& 480.1 & 436.3  & -43.8 & 7.35 & 0.96	\\
  320.0	& 640.1 & 785.8  & 145.7 & 5.45 & 1.93	\\
  480.0 & 800.1 & 1305.2 & 505.1 & 3.85 & 3.37 	\\
  \hline \end{tabular}

  \caption{Average recovered host parameters from single component
  fits to synthetic quasars with 2-component host
  galaxies. Uncorrelated Gaussian noise has been added to these models
  to match that of quasar 1543+489, and the average recovered values
  are given for 5 different realisations of this noise. The same noise
  was added to corresponding mock images created with different
  spheroidal luminosities, so there will be a systematic error because
  5 realisations are not sufficient to fully sample the recovered
  parameters with the given noise level. The result of analysing a
  host with no spheroidal component shows that the results
  systematically underestimate $\beta$ and $L_{\rm int}$ and
  overestimate $R_{1/2}$ by small amounts. Note that the relative
  dependence of the recovered parameters on the spheroidal component
  will not be affected.}  \label{tab:2cmpt}
\end{table}

Because $R_{1/2}(\rm spheroidal)$ and $L_{\rm int}(\rm spheroidal)$
follow a FP relation, the importance of this component is enhanced for
large $L_{\rm int}(\rm spheroidal)$ and diminished for small $L_{\rm
int}(\rm spheroidal)$. The recovered total luminosity for small
spheroidal components is therefore very similar to that of the disk
alone. For large spheroidal components, the modelling places an excess
of host light in the core in order to simultaneously fit the outer
disk-like profile and the inner profile with a single, large $\beta$
value. This explains the behaviour of the difference between the
actual and recovered $L_{\rm int}$ values. Recovered $\beta$
monotonically increases with the increasing luminosity of the
spheroidal core, suggesting that the spheroidal core cannot be
completely `hidden' without affecting the best fit galaxy. This adds
to the evidence that the low $\beta$ values recovered for some of the
quasars implies that they do not contain strong spheroidal
components. Note that the recovered host luminosities are
approximately correct for recovered values of $\beta$ consistent with
a host dominated by a disk-like profile.

For the quasars which have best-fit hosts dominated by spheroidal
components, a disk-like profile at larger radii could have erroneously
increased the recovered total host luminosity. However, in order to
simultaneously fit these regions, small values of $R_{1/2}$ were
required. For the quasars with hosts found to be dominated by
spheroidal components, the relatively large values of $R_{1/2}$
recovered suggest that such a disk-like component is not present.

\section{Discussion}

\subsection{Luminosities}  \label{sec:discuss:lum}

The integrated host luminosities derived from our $K$-band images
exhibit a low dispersion around a mean similar to that calculated in
studies of less luminous quasars. This is in accord with the work of
McLure \etal\ \shortcite{mclure} who also found no evidence for an
increasing trend, although they had fewer data points at high nuclear
luminosity.

Previous HST studies have found evidence that host luminosity
increases with nuclear luminosity \cite{hooper}, although the trend
observed in this work could be due to incorrect nuclear component
removal: escaping nuclear light which increases in luminosity with the
core could be added to the host light. It has recently been stated
that the psf derived by packages such as {\sc tinytim}, as used by
Hooper, Impey \& Foltz \shortcite{hooper} deviate from empirical
WFPC~2 psfs at large radii ($\ge2$ arcsec), due to scattering within
the camera \cite{mclure}, and this could be the reason for an excess
of nuclear light at larger radii which could be mistaken as host
light. This excess light could also be the reason the low axial ratios
observed in the Hooper, Impey \& Foltz \shortcite{hooper} work are not in
accord with those derived in McLure \etal\ \shortcite{mclure}, or in
this $K$-band study.

The triangular shape of the McLeod \& Rieke points in
Fig.~\ref{fig:hostvnuc} found for low redshift ($0<z<0.3$) Seyferts
and quasars of lower luminosity than those in our sample, has been
shown to be in accord with a lower limit to the host luminosity which
increases with nuclear luminosity \cite{mcleod95a}. This cut-off in
the host luminosity is equivalent to there being an upper limit to
allowed nuclear-to-host ratios. The triangular shape is {\em not}
followed by the results of the work presented in this paper which lie
to the right of the McLeod \& Rieke points. The relative positions of
the two data sets in this Figure are set by the empirical $H-K$
corrections applied to the apparent $H$-band data (see
Section~\ref{sec:calc_lum} for details). Quasar 1354+213 was included
in both our sample and the sample of McLeod \& Rieke
\shortcite{mcleod94b}, and the results of both studies independently
suggest a rest-frame nuclear-to-host ratio of $7-9$. This places
1354+213 at the right of the triangular shape of the McLeod \& Rieke
points in Fig.~\ref{fig:hostvnuc}, but it has a nuclear-to-host ratio
lower than most of the quasars in our sample, and is therefore to the
left of most of our points. We conclude that the limit suggested by
McLeod \& Rieke \shortcite{mcleod95a} must break down for quasars with
the highest nuclear luminosities.

This is in contrast to recent work by McLeod, Rieke \&
Storrie-Lombardi \shortcite{mcleod99} who claim that the lower bound
on host luminosity extends to the highest luminosity quasars, partly
based on the discovery of one luminous quasar, 1821+643 which appears
to be in a host at $\sim25L^*$. What should we expect? The hosts of
the quasars known to date already extend to about $2L^*$. Should the
hosts of quasars which are ten times more luminous be found in
galaxies at $20L^*$? Our analysis suggests not.

This result is highly important for recent quasar models. In
particular, the model of Kauffmann \& Haehnelt \shortcite{kauffmann}
predicts that the upper limit to the nuclear-to-host ratio should
extend to quasars such as those imaged in this work. However, this is
clearly not the case. A possible fix to their model would be to invoke
the scatter of the Magorrian \etal\ \shortcite{magorrian} relations to
explain high luminosity quasars (\& high mass black holes) within low
luminosity structures, and invoke a steeply-declining host mass
function to explain the lack of really massive hosts. Further work on
this model would then be required, particularly with regard to the
revised slope of the high luminosity tail of the quasar luminosity
function. Alternatively, factors other than black-hole mass, such as
nuclear obscuration, accretion processes, etc. could be the cause of
differing nuclear luminosities within reasonably similar galaxies
(with similar black hole masses).

\subsection{Morphologies}  \label{sec:discuss_morph}

Recent HST results suggest that luminous nuclear emission
predominantly arises from hosts with large spheroidal components
\cite{mclure}. The two least luminous radio-quiet quasars imaged by
McLure \etal\ \shortcite{mclure} have disk-like structure at radii
$\simgt3$\,arcsec, while the more luminous quasars are completely
dominated by spheroidal profiles. Could we be seeing a relationship
between host morphology and nuclear luminosity? This is particularly
interesting when compared to the black hole mass-spheroid mass and
spheroid mass-luminosity relations determined for nearby galaxies by
Magorrian \etal\ \shortcite{magorrian}: a large black hole,
potentially capable of powering luminous AGN appears more likely to be
present in galaxies with large spheroidal components. Both the results
of McLure \etal\ \shortcite{mclure} and the relations of Magorrian
\etal\ \shortcite{magorrian} suggest that quasars with strong nuclear
emission should predominantly exist in hosts with large spheroidal
components which dominate any disk-like structure.

By careful analysis we have provided evidence that a large fraction of
the host galaxies found in this work are dominated by disk-like
profiles. However, the most important light for this modelling comes
from radii greater than those of the HST study, where the disk
component, if present, is expected to be strong. The $K$-band images
described here are not of sufficient quality for us to resolve the
inner region and produce a 2-component fit to the host galaxy. This is
in contrast to results from HST where the increased resolution enables
the inner region to be resolved, and the spheroidal core of the galaxy
becomes more important for modelling with a single $\beta$
parameter. By analysing synthetic data, we have been able to show that
for hosts where we find a dominant disk-like component, any additional
spheroidal component will not result in a large change in the
recovered total host luminosities. We have also provided suggestive
evidence that the spheroidal cores of these quasars are of relatively
low luminosity. Further analysis of both the regions and profiles
probed by different studies, and higher resolution data on the cores
of the quasars analysed in this work would be very interesting, and
could help to explain the different morphological results of recent
host galaxy studies.

\section{Conclusions}

We have presented the results from a deep $K$-band imaging study
designed to reveal the host galaxies of quasars with higher
luminosities than targeted by previous studies. Extending host-galaxy
studies to these quasars was made possible by the stability provided
by the tip-tilt adaptive optics system at UKIRT, which enabled
accurate psf measurements to be obtained for the deep quasar
images. We have been able to resolve host galaxies for all of our
sample.

The principle conclusion of this study is that the luminous quasars in
this sample have host galaxies with similar luminosities to quasars of
lower total luminosity. Derived nuclear-to-host ratios are therefore
larger than those obtained by previous work, and place these quasars
beyond the upper limit suggested by studies of quasars with lower
total luminosities. Host morphologies are less certain, but there is
weak evidence that the hosts of these quasars can be dominated by
either disk-like or spheroidal profiles on the scales probed by
these images.

\section{Acknowledgements}
The United Kingdom Infrared Telescope is operated by the Joint
Astronomy Centre on behalf of the U.K. Particle Physics and Astronomy
Research Council.

\end{document}